\documentclass[journal]{IEEEtran}
\usepackage{amsfonts,amssymb}
\usepackage[fleqn]{amsmath}
\usepackage{mathtools}
\usepackage{dsfont}
\usepackage{bm}
\usepackage{stmaryrd}
\usepackage{makecell}
\usepackage{empheq}
\usepackage{algorithm}
\usepackage{algpseudocode}
\usepackage{array}
\usepackage{textcomp}
\usepackage{stfloats}
\usepackage{url}

\usepackage{verbatim}
\usepackage[dvipsnames]{xcolor}
\usepackage{graphicx}
\def\BibTeX{{\rm B\kern-.05em{\sc i\kern-.025em b}\kern-.08em
    T\kern-.1667em\lower.7ex\hbox{E}\kern-.125emX}}
\usepackage{balance}
\usepackage{hyperref}
\hypersetup{
  colorlinks=true,
  linkcolor=red,
  filecolor=magenta,
  urlcolor=cyan,
  citecolor=blue,
}
\usepackage{cleveref}
\usepackage{booktabs}
\usepackage{multirow}
\usepackage{arydshln}
\usepackage{tikz}
\usepackage{pgfplots}
\usepgfplotslibrary{groupplots}
\pgfplotsset{compat=1.18}

\usepackage{lipsum}
\usepackage{transparent}

\usepackage{subcaption}

\algrenewcommand\algorithmicindent{.6em} 

\newcommand{\bTheta}{\bm{\Theta}}
\newcommand{\btheta}{\bm{\theta}}
\newcommand{\bY}{\bm{Y}}
\newcommand{\bU}{\bm{U}}
\newcommand{\by}{\bm{y}}
\newcommand{\bu}{\bm{u}}
\newcommand{\nl}{_{n,\ell}}
\newcommand{\nd}{_{n,d}}
\newcommand{\mcite}[1][]{\begingroup\color{red}[Citation needed\if\relax\detokenize{#1}\relax\else: #1\fi]\endgroup}

\pgfplotsset{
    my common style/.style={
        scale only axis,
        width=.9\linewidth, 
        height=2.5cm,         % Hauteur stricte de la zone de dessin
        samples=400, % High sample count for smooth curves
        legend style={
            font=\small, rounded corners=2pt, fill=white,
            fill opacity=0.7, text opacity=1, draw={gray!40},
            align=left, legend cell align=left
        },
        legend pos=north east,
        tick style={major tick length=2pt, minor tick length=1pt, thin},
        tick label style={font=\footnotesize},
        minor tick num=1,
        grid=major,
        major grid style={line width=0.5pt,draw=gray!40},
        extra x tick style={
            grid=none, 
            tick style={draw=none}, 
            tick label style={font=\footnotesize\bfseries, yshift=0pt}
        }
    }
}

\definecolor{mycyan}{HTML}{0AE85F}   % bright cyan/teal
\definecolor{mypurple}{HTML}{7E1CEE} % purple
\definecolor{myorange}{HTML}{FF9E37} % orange

\begin{document}
\bstctlcite{IEEEexample:BSTcontrol}
\title{A Hierarchical Model for Non-linear Inverse Problems under Additive and Multiplicative Noise}

\author{
    \IEEEauthorblockN{Nicolas GOEMAN, Pierre-Antoine THOUVENIN, Pierre CHAINAIS}\\
    \IEEEauthorblockA{Univ. Lille, CNRS, Centrale Lille, UMR 9189 CRIStAL, F-59000 Lille, France}%
    \thanks{This work was supported by the ANR project ``Chaire IA Sherlock'' ANR-20-CHIA-0031-01 hold by P. Chainais, as well as by the national support within the programme d'investissements d'avenir ANR-16-IDEX-0004 ULNE and R\'egion HDF. 
    HPC and storage resources were provided by the HPC Mésocentre at the University of Lille.}%
}

\markboth{Journal of \LaTeX\ Class Files,~Vol.~18, No.~9, September~2020}%
{How to Use the IEEEtran \LaTeX \ Templates}

\maketitle

\begin{abstract}
Ill-posed inverse problems are encountered in numerous applications, possibly characterized by a highly non-linear forward model, both additive and multiplicative sources of noise, and censored data. 
In the absence of ground truth, uncertainty quantification is crucial to assess estimation reliability. This motivates the use of a Bayesian model and stochastic inference methods such as Markov Chain Monte Carlo algorithms. Problems combining all these challenges often lead to a complex and potentially multimodal posterior distribution, difficult to handle in practice. 
Approximate approaches have been proposed in the literature by either neglecting a source of noise or by using a tractable approximation of the likelihood function. 
These approaches either lead to an inaccurate model, or may require a complex calibration of the approximate likelihood.
This paper proposes to tackle such problems with a general hierarchical Bayesian model and an efficient MCMC algorithm. 
The proposed formulation bypasses the need for calibrating the hyperparameters of an approximate model and is more versatile.
The proposed method is assessed on a challenging scenario encountered in astronomy using synthetic data, in a variety of noise and censoring configurations.
Comparisons are conducted against two baselines and a state-of-the-art method applicable in this context.
The proposed approach is general and yields state-of-the-art results in terms of point-wise estimates and computing costs, with superior predictive performance.
Results in the supplementary material further complete this comprehensive and rigorous model study. This work can serve as a guide for practitioners to select the best likelihood model according to their specific application.
\end{abstract}


\section{Introduction}
\label{sec:intro}

This paper focuses on a family of non-linear inverse problems using censored observations subject to both additive and multiplicative sources of noise.
Moreover, in many scenarios, both the input and output of the non-linear forward operator may have high dynamics and span several decades in amplitude. The additive and multiplicative sources of noise therefore need to be faithfully accounted for, since the true nature of the dominant noise strongly varies with amplitudes. %Indeed, the nature of the dominant noise can be hard to predict a priori, since the contribution of the multiplicative noise depends on the amplitude of the forward operator output.

Such problems are encountered in several applications, for example in astronomy \cite{Kelly_Shetty_Stutz_Kauffmann_Goodman_Launhardt_2012,Palud_Bron_Chainais_Petit_Thouvenin_Santa-Maria_Goicoechea_Languignon_Gerin_Pety_et_al_2025}, ultrasound imaging \cite{Krissian_Westin_Kikinis_Vosburgh_2007} or SAR imaging \cite{Durand_Fadili_Nikolova_2008}.
In practice, ground truth data is rarely available to calibrate model hyperparameters and assess estimation quality.
These limitations necessitate robust uncertainty quantification, motivating the use of Bayesian methods.

The objective is to infer physical parameters $\boldsymbol{\Phi} = (\boldsymbol{\phi}_1, \ldots, \boldsymbol{\phi}_N)^{\intercal} \in \mathbb{R}^{N \times D}$ from observations $\bm{Y}=(y\nl)\nl=(\bm{y}_1, \ldots, \bm{y}_N)^{\intercal} \in \mathbb{R}^{N \times L}$, where $N$ is the number of observations $\bm{y}_n \in \mathbb{R}^L$ and $L$ is the dimensionality of the observations. In the remaining of this paper, the dimension $N$ is interpreted as the number of pixels in a multispectral map, and $L$ the number of spectral bands.
Each observation $y\nl$ is related to the $D$ parameters $\boldsymbol{\phi}_n$ through a deterministic forward operator $\bm{g} \colon \boldsymbol{\phi}_n \in \mathbb{R}^D \mapsto \bm{g}(\boldsymbol{\phi}_n) = (g_\ell(\boldsymbol{\phi}_n))_{1 \leq \ell \leq L} \in \mathbb{R}_+^L$, and noise.
This paper considers the following model, for any $(n,\ell) \in \llbracket 1, N \rrbracket \times \llbracket 1, L \rrbracket$,
\begin{equation}
  \label{eq:obs-model}
  y\nl = \max\bigl\{\omega\nl, \; \xi\nl\;g_\ell(\boldsymbol{\phi}_n) \; + \; \epsilon\nl\bigr\},
\end{equation}
where $\omega\nl \in \mathbb{R}$ is a threshold representing the limited sensitivity of the observational instrument, $\epsilon\nl \sim \mathcal{N}(0, \;\sigma_{a,n,\ell}^2)$ is an additive noise, $\xi\nl \sim \textrm{Log}\mathcal{N}(-\sigma_m^2/2,\; \sigma_m^2)$ is a multiplicative noise independent of $\epsilon\nl$ such that $\mathbb{E}[\xi\nl] = 1$, and $\sigma_{a,n,\ell}^2, \sigma_m^2 > 0$.

The likelihood of model~\eqref{eq:obs-model} is intractable, and more difficult to handle than for typical settings with a single source of noise~\cite{Krissian_Westin_Kikinis_Vosburgh_2007,Durand_Fadili_Nikolova_2008,Shi_Osher_2008}. 
%Two main approaches have been considered in the literature to deal with this issue.
%
A first approach neglects one source of noise when the other is dominant~\cite{Krissian_Westin_Kikinis_Vosburgh_2007,Durand_Fadili_Nikolova_2008}, leading to a cheaper and tractable likelihood.
When the multiplicative noise dominates, it can be recast into an additive noise by taking the logarithm of the observation model~\cite{Shi_Osher_2008}.
When both sources of noise play a significant role, in particular when the forward model covers a high dynamic range, the likelihood must be adapted.
In~\cite{Nicholson_Kaipio_2020}, the two sources of noise are taken into account through the parameterization of the covariance matrix of a Gaussian model, computed using a Bayesian approximation error method.
This approach scales well to high dimensional problems, as the posterior distribution remains Gaussian for classical choices of prior distribution.
In \cite{Huang_Ng_Zeng_2013}, the multiplicative noise is treated explicitly alongside the additive noise with a hierarchical model.
The resulting non-convex optimization problem is relaxed into a more tractable version to form a \emph{maximum a posteriori} (MAP) estimate using an iterative algorithm.
Note that these works \cite{Nicholson_Kaipio_2020,Huang_Ng_Zeng_2013} still propose approximate likelihoods, for linear forward models only.
% Palud's work
To the best of the authors' knowledge, \cite{Palud_Thouvenin_Chainais_Bron_LePetit_2023} is the only work in the literature that explicitly addresses model \eqref{eq:obs-model} with all together non-linearity, high dynamic ranges, additive and multiplicative noises and censoring.
Moreover, the forward operator can be encoded by a highly non-linear black-box model.
The authors of~\cite{Palud_Thouvenin_Chainais_Bron_LePetit_2023} propose an approximate Bayesian model and a Markov Chain Monte Carlo (MCMC) algorithm to draw samples from the resulting posterior distribution.
They propose a likelihood approximation based on a smooth interpolation between a purely additive and a purely multiplicative noise model.
For low amplitudes of the forward operator, the additive noise dominates the multiplicative noise, and conversely for high amplitudes. 
The transition between these two regimes is controlled by a sigmoid-like weighting function. 
The beginning and end of the transition are critical hyperparameters. 
They need to be carefully fine-tuned for each application and level of noise. 
In addition, the resulting approximate likelihood has an intractable normalization constant depending on $\boldsymbol{\Phi}$ that needs to be estimated.
A major caveat of all these prior works is that the bias introduced by these approximations has not been assessed in a comprehensive study. 

To address these limitations, we propose a reformulation of \eqref{eq:obs-model} based on a general hierarchical model leading to a tractable likelihood  that exactly accounts for both sources of noise.
This avoids the need for a dedicated approximation and complex hyperparameter calibration.
For inference, we propose an MCMC algorithm inspired by~\cite{Palud_Thouvenin_Chainais_Bron_LePetit_2023} to efficiently sample the resulting posterior distribution and build the estimators of interest.
Finally, a rigorous and careful comparison of different approximate likelihood  models with the proposed hierarchical model is conducted. 
It permits to better understand the impact of likelihood models on the estimation quality, both in terms of parameter reconstruction quality and uncertainty quantification. 
This study provides a guide for practitioners to select an appropriate likelihood model in similar settings, based on their specific context of application.

The remaining of the paper is organized as follows.
\Cref{sec:model} describes the hierarchical formulation considered for the likelihood, the prior distribution and the resulting posterior distribution. 
\Cref{sec:sampler} details the proposed MCMC sampler and the associated proposal distributions. 
\Cref{sec:experiments} reports numerical experiments on synthetic astronomical data to demonstrate the performance of the proposed method. 
\Cref{sec:conclusion} summarizes the contributions of this work and outlines research perspectives.
\section{Proposed Bayesian hierarchical model}
\label{sec:model}

\subsection{High dynamic range reparameterization}

The physical parameters of interest $\boldsymbol{\Phi}$ can span several orders of magnitude. 
This might lead to numerical instabilities and difficulties at inference time. 
To this end, the sampled parameters are not the physical parameters $\boldsymbol{\Phi}$ themselves but a reparametrization $\bTheta  = (\btheta_1, \ldots, \btheta_N)^{\intercal} \in \mathbb{R}^{N \times D}$ encoded by a elementwise transformation $\boldsymbol{\nu} \colon \mathbb{R}^D \rightarrow \mathbb{R}^D$.
This transformation operates element-wise such that $\boldsymbol{\nu}(\boldsymbol{\phi}_n) = (\nu_1(\phi_{n,1}), \ldots, \nu_D(\phi_{n,D}))$, where each scalar function $\nu_d \colon \mathbb{R} \rightarrow \mathbb{R}$ is defined for $d\in \llbracket 1, D\rrbracket$ as
\begin{equation}
  \label{eq:normalized_space}
  \theta_{n,d} = \nu_d(\phi_{n,d}) \coloneq \frac{\log_{10}(\phi_{n,d}) - \mu_d}{\sigma_d},
\end{equation}
% \ngoe{[$\mu_d$ and $\sigma_d$ kind of pop out of nowhere without speaking of $f$ ?]}
where $\mu_d$ and $\sigma_d$ are the empirical mean and standard deviation of a predefined grid of points covering the physical parameter space~\cite{Palud_Thouvenin_Chainais_Bron_LePetit_2023}, e.g., when $\boldsymbol{g}$ is encoded by a pre-trained neural network.
Using this reparameterization, the forward operator becomes $\boldsymbol{f} = \boldsymbol{g} \circ \boldsymbol{\nu}^{-1}$.

In many physical applications, the forward operator is encoded by a computationally-intensive non-linear numerical model \cite{Doicu_Efremenko_Wirth_Wriedt_2025,Dinkel_Geitner_Rei_Nitzler_Wall_2023}. It is then often replaced by a cheaper surrogate model $\boldsymbol{f}$, \emph{e.g.}, a Gaussian process~\cite{Rasmussen_2003} or a neural network~\cite{Yan_Zhou_2020}.

\subsection{Hierarchical likelihood}

We consider a two stage hierarchical reformulation of \eqref{eq:obs-model} that introduces an auxiliary variable.
For any $(n, \ell) \in \llbracket 1, N \rrbracket \times \llbracket 1, L \rrbracket$, the model \eqref{eq:obs-model} can be equivalently rewritten
\begin{subequations}
\label{eq:hier-model}
\begin{empheq}[left=\empheqlbrace]{align}
    y_{n,\ell} &= \max \bigl\{ \omega_{n,\ell},\: u_{n,\ell} + \epsilon_{n,\ell} \bigr\} \label{eq:hier-model-u2}\\
    u_{n,\ell} &= \xi_{n,\ell} \, f_{\ell}(\boldsymbol{\theta}_n), \label{eq:hier-model-u1}
\end{empheq}
\end{subequations}
with $\xi_{n,\ell} \sim \log\mathcal{N}(-\sigma_m^2/2, \sigma_m^2)$ and $\epsilon_{n,\ell} \sim \mathcal{N}(0, \sigma_{a,n,\ell}^2)$.
The auxiliary variable $u\nl$ is intended to decouple the impact of the noise sources on the parameters $\btheta_n$.

Assuming conditional independence across the pixels $n$ and spectral bands $\ell$, the likelihood function then factorizes into
\begin{equation}
  \label{eq:likelihood}
\pi(\boldsymbol{Y}, \boldsymbol{U} \mid \bTheta) =  \prod_{n=1}^{N}\prod_{\ell=1}^{L} \pi(y\nl\vert u\nl) \: \pi(u\nl\vert \btheta_n),
\end{equation}
where $\pi(u\nl\vert \btheta_n)$ is the probability density function (pdf) of the distribution $\log\mathcal{N}\Bigl(u\nl \ ; \ \log\bigl(f_{\ell}(\boldsymbol{\theta}_n)\bigr) - \sigma^2_m/2, \: \sigma^2_m\Bigr)$ from the multiplicative noise term \eqref{eq:hier-model-u1}. 
Due to censoring, the term $\pi(y\nl\vert u\nl)$ is the pdf of a generalized rectified Gaussian distribution \cite{Palmer_Hill_Scheding_2017} with lower bound $\omega\nl$ \eqref{eq:hier-model-u2}:
\begin{equation}
    \begin{split}    
        \pi(y\nl\vert u\nl) =\ &\phi(y\nl \vert u\nl, \sigma_{a,n,\ell}^2) \: \mathds{1}_{(\omega\nl, +\infty)}(y\nl) \\
        &+ \Phi(\omega\nl \vert u\nl, \sigma_{a,n,\ell}^2) \: \delta(y\nl-\omega\nl),
    \end{split}
\end{equation}
with $\phi(\cdot \vert \mu, \sigma^2)$ the pdf of a Gaussian distribution with mean $\mu$ and variance $\sigma^2$, $\Phi(\cdot \vert \mu, \sigma^2)$ its cumulative distribution function (cdf), $\mathds{1}_{A}$ the characteristic function of the set $A$ and $\delta$ the Dirac distribution. 

The hierarchical formulation~\eqref{eq:hier-model} leads to a more convenient  expression of the likelihood function for inference, while statistically equivalent to~\eqref{eq:obs-model}. 
It avoids the need for an approximate likelihood model with additional hyperparameters to tune. 
Besides, the two main challenges in~\eqref{eq:hier-model}, \emph{i.e.}, the highly non-linear forward operator on one hand and the mixture of two sources of noise combined with censoring on the other hand, only affect a single variable at once.

\begin{algorithm}[t]
\caption{Global sampling procedure}
\label{alg:global_sampler}
\begin{algorithmic}[1]
\State \textbf{Input:} $p \in [0,1]$, $T$, $T_{BI}$, $G=(G_1, G_2)$
\State \textbf{Initialize:} $\bm{\Theta}^{(0)}$, $\bm{U}^{(0)}$
\For{$t = 0$ to $T-1$}
  \State Sample $z^{(t+1)} \sim \mathcal{U}(0,1)$ \Comment{Kernel selection}
  \For{$g \in G$} \Comment{Independent index set}
    \For{$n \in g$} \Comment{In parallel}
      \If{$z^{(t+1)} < p$}
        \State Draw $(\bm{\theta}_n^{(t+1)}, \bm{u}_n^{(t+1)})$ using MH (see \Cref{sec:kernels_theta})
      \Else
        \State Draw $(\bm{\theta}_n^{(t+1)}, \bm{u}_n^{(t+1)})$ using MTM (see \Cref{sec:kernels_theta})
      \EndIf
      \State Update PMALA kernel parameters (see \cite{Palud_Thouvenin_Chainais_Bron_LePetit_2023})
    \EndFor
  \EndFor
\EndFor
\State \textbf{Output:} Posterior samples $\{\bm{\Theta}^{(t)}, \bm{U}^{(t)}\}_{t=T_{BI}+1}^T$
\end{algorithmic}
\end{algorithm}

\subsection{Prior and posterior distributions}

We consider the same prior distribution on $\bTheta$ as in \cite{Palud_Thouvenin_Chainais_Bron_LePetit_2023}, based on two smooth regularization functions.
The use of differentiable functions allows gradient-based transition kernels to be used for an efficient MCMC algorithm.

The first regularization function $h \colon \mathbb{R}^N \to \mathbb{R}_+$ promotes spatially-smooth parameter maps, a common requirement for several applications in physics.
A squared $\ell_2$-norm is considered to penalize the discrete gradient~\cite{Calvetti_Somersalo_2018} of each parameter map $\bTheta_{\cdot, d} = (\theta_{n,d})_{n=1}^N$, leading to
\begin{equation}
  \label{eq:spatial_prior}
  h(\bTheta_{\cdot d}) = \frac{1}{2}\sum_{n=1}^{N} \sum_{i \in \mathcal{V}_n} (\theta_{n,d} - \theta_{i,d})^2,
\end{equation} 
where $\mathcal{V}_n$ is the index set of neighboring pixels of pixel $n$. 
The second regularizer constrains $\bTheta$ to remain within a compact set $\mathcal{C} = [l_1, u_1] \times \ldots \times [l_D, u_D]$, ensuring physical consistency.
As in \cite{Palud_Thouvenin_Chainais_Bron_LePetit_2023}, for each $d \in \llbracket 1, D\rrbracket$, a smooth approximation to the indicator function $\iota_{\mathcal{C}_d}$\footnote{The indicator function $\iota_{\mathcal{C}}$ of a non-empty set $\mathcal{C}$ is defined by $\iota_{\mathcal{C}}(\bm{x}) = 0$ if $\bm{x} \in \mathcal{C}$, and $+\infty$ otherwise.} is used.
This approximation $\widetilde{\iota}_{\mathcal{C}_d} \colon \mathbb{R} \to \mathbb{R}_+$ is defined by
\begin{equation}
  \label{eq:smooth_indicator}
  \widetilde{\iota}_{\mathcal{C}_d}(\theta_{n, d}) = \max\{0, \theta_{n,d}-u_d, l_d-\theta_{n,d}\}^4.
\end{equation}
The resulting prior distribution reads
\begin{equation}
  \label{eq:prior}
  \pi(\boldsymbol{\Theta}) \quad \propto \quad \prod_{d=1}^{D} \exp\biggl[ -\tau_d h\bigl(\bTheta_{\cdot d}\bigr) - \delta \sum_{n=1}^{N}\widetilde{\iota}_{\mathcal{C}_d}(\theta_{n, d}) \biggr],
\end{equation}
where $\boldsymbol{\tau} = (\tau_d)_{1 \leq d \leq D} \in \mathbb{R}_+^D$ contains the spatial regularization hyperparameters, and $\delta \in \mathbb{R}_+$ is the hyperparameter associated to the validity constraint.
The vector $\boldsymbol{\tau}$ controls the spatial smoothness enforced on the parameter maps.

Combining the hierarchical likelihood \eqref{eq:likelihood} with the prior distribution \eqref{eq:prior} yields the posterior distribution pdf
\begin{equation}
    \pi(\bTheta, \bU \mid \bY) \propto  \pi(\bY, \bU \mid \bTheta) \: \pi(\bTheta).
    \label{eq:posterior}
\end{equation}

\section{Proposed MCMC sampler}
\label{sec:sampler}

This section describes the proposed MCMC sampler for the model defined in \Cref{sec:model}. 
The factorization of~\eqref{eq:posterior} suggests that groups of variables can be updated at once, conditionally upon others. 
To this end, a conditional Metropolis-Hastings \cite{Jones_Roberts_Rosenthal_2014} -- also called Metropolis-within-Gibbs sampler -- is used to update $\bTheta$ and $\bU$, ensuring convergence under mild conditions.

\subsection{Block updates and global sampling strategy}

To define an efficient global sampling strategy, $\btheta_n$ and $\bu_n$ are grouped into a single block update to account for their strong coupling induced by the pdf $\pi(u\nl \mid \btheta_n)$ in \eqref{eq:likelihood}.
Indeed, proposing a candidate $\tilde{\btheta}_n^{(t+1)}$ in a new mode of the posterior distribution given $\bu_n^{(t)}$ can be unlikely with respect to $\pi(u\nl^{(t)} \mid \tilde{\btheta}_n^{(t+1)})$ when the discrepancy between $f(\tilde{\btheta}_n^{(t+1)})$ and $u\nl^{(t)}$ is too large. 
The same situation occurs when proposing a new candidate $\tilde{\bu}_n^{(t+1)}$ given $\btheta_n^{(t)}$. 
The accept/reject step of $(\btheta_n, \bu_n)$ should therefore be coordinated to efficiently explore different modes in the parameter space and avoid low acceptance rates.

The spatial and spectral factorization of both the hierarchical likelihood (\ref{eq:likelihood}) and part of the prior~\eqref{eq:smooth_indicator}-\eqref{eq:prior} further motivates a Gibbs-based approach to update the parameters $(\btheta_n, \bu_n)$. Sampling from the posterior \eqref{eq:posterior} therefore amounts to sampling from the $N$ full conditional distributions, with pdf given for $n \in \llbracket 1, N \rrbracket$  by
\begin{equation}
\begin{split}
  \pi(\btheta_n, &\bu_n \mid \by_n, \bTheta_{\backslash n}, \, \bU_{\backslash n}) \; = \; \pi(\btheta_n, \bu_n \mid \by_n, \bTheta_{\mathcal{V}_n})\\
  &\propto \; \pi(\btheta_n \mid \bTheta_{\mathcal{V}_n}) \: \prod_{\ell=1}^{L} \: \pi(u\nl \mid \btheta_n) \: \pi(y\nl \mid u\nl),
  \label{eq:conditional_posterior}
\end{split}
\end{equation}
where $\bTheta_{\backslash n}$ and $\bU_{\backslash n}$ denotes the set of all parameter maps without the $n$-th pixel. 
The neighboring structure of the spatial regularization \eqref{eq:spatial_prior} partitions the maps into two groups $G_1$ and $G_2$ following a checkerboard pattern. Pixels in a group are independent conditionally on those in the other group. This allows all pixels in a group to be updated in parallel in a chromatic Gibbs sampling approach \cite{Gonzalez_Low_Gretton_Guestrin_2011}, which significantly speeds up the algorithm.

Direct sampling from the full conditional distribution \eqref{eq:conditional_posterior} is intractable. A joint proposal distribution for the pair $(\boldsymbol{\theta}_n, \boldsymbol{u}_n)$ is therefore considered within an appropriate transition kernel. 
The proposal mirrors the hierarchy in \eqref{eq:conditional_posterior} and factorizes as
\begin{equation}
\begin{split}
    \label{eq:proposal_total}
    q(\boldsymbol{\theta}_n, \boldsymbol{u}_n \mid \boldsymbol{\theta}_n^{(t)}, \boldsymbol{u}_n^{(t)}) &\propto q_{\boldsymbol{\theta}}(\boldsymbol{\theta}_n \mid \boldsymbol{\theta}_n^{(t)}, \boldsymbol{u}_n^{(t)})\\
    &\times \ \prod_{\ell=1}^{L}q_{u}(u_{n,\ell} \mid \boldsymbol{\theta}_n, y_{n,\ell}).
\end{split}
\end{equation}
In other words, a new candidate state $\tilde{\btheta}_n^{(t+1)}$ is first proposed for $\btheta_n$, and then for every $u\nl$, $\ell \in \llbracket 1, L \rrbracket$, conditionally on $\tilde{\btheta}_n^{(t+1)}$. The couple $(\tilde{\btheta}_n^{(t+1)}, \tilde{\boldsymbol{u}}^{(t+1)})$ is then jointly accepted or rejected according to an acceptance ratio.

We address the potential multimodality and non-linearity of the posterior distribution with a mixture of 2 transition kernels inspired by~\cite{Palud_Thouvenin_Chainais_Bron_LePetit_2023}.
The first kernel $K^{(1)}$ uses a local gradient-based proposal inside a Metropolis-Hastings (MH) algorithm~\cite{Robert_Casella_2004}. 
The second kernel $K^{(2)}$ is a Multiple-Try Metropolis (MTM) algorithm~\cite{Liu_Liang_Wong_2000} designed for global exploration \emph{via} a long-range proposal distribution. 
The sampler is a random mixture of these two kernels, with probability $p \in (0, 1)$ for the MH scheme and $1-p$ for the MTM scheme. The resulting transition kernel $K$ is
\begin{equation}
\begin{split}
  K(\btheta_n, \bu_{n} \mid &\btheta_n^{(t)}, \bu_n^{(t)}) =\; p \: K^{(1)}(\btheta_n, \bu_n \mid \btheta_n^{(t)}, \bu_n^{(t)}) \\
  &+ (1-p) \: K^{(2)}(\btheta_n, \bu_n \mid \btheta_n^{(t)}, \bu_n^{(t)}).
\end{split}
\end{equation}
\Cref{alg:global_sampler} summarizes the structure of the proposed sampler.

\Cref{sec:kernels_theta} details the respective proposal distributions $q_{\btheta}^{(1)}$ and $q_{\btheta}^{(2)}$ of the two kernels $K^{(1)}$ and $K^{(2)}$. 
\Cref{sec:proposal_u} describes the proposal distribution $q_u$ used in both kernels $K^{(1)}$ and $K^{(2)}$ for the auxiliary variables $u\nl$.

\subsection{Kernels and proposal distributions for \texorpdfstring{$\btheta_n$}{theta_n}}
\label{sec:kernels_theta}

\paragraph{Local exploration kernel $K^{(1)}$}

The proposal distribution $q_{\btheta}^{(1)}$ used for local exploration is based on a position-dependent preconditioned Metropolis-adjusted Langevin algorithm (PMALA) \cite{Xifara_Sherlock_Livingstone_Byrne_Girolami_2014,Roberts_Tweedie_1996}, designed as in \cite{Palud_Thouvenin_Chainais_Bron_LePetit_2023}. For completeness, the proposal is recalled in the supplementary material.
The candidate couple $(\tilde{\btheta}_n^{(t+1)}, \tilde{\bu}_n^{(t+1)})$ is then accepted with the MH acceptance probability $a^{(t+1)} = \min\{1, r^{(t+1)}\}$, with
\begin{equation}
  \label{eq:acceptance_ratio_mh}
  \begin{aligned}
    r^{(t+1)} &= \frac{\pi(\tilde{\btheta}_n^{(t+1)}, \tilde{\bu}_n^{(t+1)} \mid \by_n, \bTheta_{\backslash n}^{(t)})}{\pi(\btheta_n^{(t)}, \bu_n^{(t)} \mid \by_n, \bTheta_{\backslash n}^{(t)})} \\
    &\qquad \times \frac{q^{(1)}(\btheta_n^{(t)}, \bu_n^{(t)} \mid \tilde{\btheta}_n^{(t+1)}, \tilde{\bu}_n^{(t+1)})}{q^{(1)}(\tilde{\btheta}_n^{(t+1)}, \tilde{\bu}_n^{(t+1)} \mid \btheta_n^{(t)}, \bu_n^{(t)})}.
  \end{aligned}
\end{equation}
\Cref{alg:mh} summarizes the PMALA-based kernel $K^{(1)}$.

\begin{algorithm}[t]
  \caption{PMALA-based kernel $K^{(1)}$}
  \label{alg:mh}
  \begin{algorithmic}[1]
  \State \textbf{Input:} Current state $(\bm{\theta}_n^{(t)}, \bm{u}_n^{(t)})$, $\bTheta_{\backslash n}^{(t)}$
  \State $\tilde{\bm{\theta}}_n^{(t+1)} \sim q_{\btheta}^{(1)}(\tilde{\bm{\theta}}_n^{(t+1)} \mid \btheta_n^{(t)}, \bm{u}_n^{(t)}, \bTheta_{\backslash n}^{(t)})$ \Comment{PMALA proposal}
  \For{$\ell = 1$ to $L$} \Comment{In parallel}
    \State $\tilde{u}_{n,\ell}^{(t+1)} \sim q_u(\tilde{u}_{n,\ell}^{(t+1)} \mid \tilde{\btheta}_n^{(t+1)})$ \Comment{See \Cref{sec:proposal_u}}
  \EndFor
  \State Compute acceptance probability $a^{(t+1)}$ \Comment{See \eqref{eq:acceptance_ratio_mh}}
  \State Sample $z^{(t+1)} \sim \mathcal{U}(0,1)$
  \If{$z^{(t+1)} < a^{(t+1)}$}
    \State $(\bm{\theta}_n^{(t+1)}, \bm{u}_n^{(t+1)}) = (\tilde{\bm{\theta}}_n^{(t+1)}, \tilde{\bm{u}}_n^{(t+1)})$
  \Else
    \State $(\bm{\theta}_n^{(t+1)}, \bm{u}_n^{(t+1)}) = (\bm{\theta}_n^{(t)}, \bm{u}_n^{(t)})$
  \EndIf
  \end{algorithmic}
\end{algorithm}

\paragraph{Global exploration kernel $K^{(2)}$}
The proposal $q^{(2)}_{\boldsymbol{\theta}}$ in the MTM-based kernel $K^{(2)}$ aims at escaping local modes explored by $K^{(1)}$. 
MTM proposes $M$ candidates at each iteration and accept one of them based on importance weights.
See~\cite{Martino_2018} for more details on MTM methods and the Independent MTM (I-MTM) variant used in this work.
The proposal distribution $q_{\btheta}^{(2)}(\:\cdot \mid \bTheta^{(t)}_{\backslash n})$ is based on the spatial regularization, see \Cref{sec:model}.
In the spirit of \cite{Palud_Thouvenin_Chainais_Bron_LePetit_2023}, $q_{\boldsymbol{\theta}}^{(2)}$ is defined as a mixture of Gaussian distributions. Each component is centered on the mean of a non-empty subset $V$ of $\mathcal{V}_n$ with precision $\tau_d \lvert V \rvert$:
\begin{multline}
  q_{\btheta}^{(2)}(\btheta_n \mid \bTheta_{\backslash n}^{(t)}) \propto \prod_{d=1}^{D}\sum_{V\in \mathcal{P}(\mathcal{V}_n)\backslash\emptyset}\biggl[ \sqrt{\tau_d\lvert V\rvert} \\
  \exp\bigl(-\tau_d\lvert V\rvert(\theta_{n,d} - \frac{1}{\lvert V\rvert}\sum_{i\in V} \theta_{i,d}^{(t)})^2\bigr)\biggr],
\end{multline}
where $\lvert V \rvert$ is the cardinality of the index set $V$. In contrast with \cite{Palud_Thouvenin_Chainais_Bron_LePetit_2023}, contributions of the modes are reweighted to have equal probabilities. A set of $M$ candidates $\{(\tilde{\boldsymbol{\theta}}_n^{(t+1)}, \tilde{\boldsymbol{u}}_n^{(t+1)})[m]\}_{m=1}^{M}$ is generated from the proposal distribution $q_{\boldsymbol{\theta}}^{(2)}$ and $q_u$. The selection of one candidate couple $(\tilde{\btheta}_n^{(t+1)}, \tilde{\bu}_n^{(t+1)})[i]$, $i \in \llbracket 1, M \rrbracket$, and its probability of acceptance depends on the standard weight function for I-MTM \cite[Section 4.1.1, p.8]{Martino_2018}. 
\Cref{alg:mtm} summarizes the I-MTM kernel $K^{(2)}$.

\begin{algorithm}[t]
\caption{I-MTM based kernel $K^{(2)}$}
\label{alg:mtm}
\begin{algorithmic}[1]
\State \textbf{Input:} Current state $(\bm{\theta}_n^{(t)}, \bm{u}_n^{(t)})$, $\bTheta_{\backslash n}^{(t)}$
\For{$m = 1$ to $M$} \Comment{In parallel}
  \State $\tilde{\bm{\theta}}_n^{(t+1)}[m] \sim q_{\btheta}^{(2)}(\tilde{\bm{\theta}}_n^{(t+1)}[m] \mid \bTheta_{\backslash n}^{(t)})$
  \For{$\ell = 1$ to $L$} \Comment{In parallel}
    \State $\tilde{u}_{n,\ell}^{(t+1)}[m] \sim q_u(\tilde{u}_{n,\ell}^{(t+1)}[m] \mid \tilde{\bm{\theta}}_n^{(t+1)}[m])$
  \EndFor
\EndFor
\State Select $(\tilde{\bm{\theta}}_n^{(t+1)}, \tilde{\bm{u}}_n^{(t+1)})[i]$ among $\{(\tilde{\bm{\theta}}_n^{(t+1)}, \tilde{\bm{u}}_n^{(t+1)})[m]\}_{m=1}^M$ (see \cite{Martino_2018})
\State Compute acceptance probability $a^{(t+1)}$ (see \cite{Martino_2018})
\State Sample $z^{(t+1)} \sim \mathcal{U}(0,1)$
\If{$z^{(t+1)} < a^{(t+1)}$}
  \State $(\bm{\theta}_n^{(t+1)}, \bm{u}_n^{(t+1)}) = (\tilde{\bm{\theta}}_n^{(t+1)}, \tilde{\bm{u}}_n^{(t+1)})[i]$
\Else
  \State $(\bm{\theta}_n^{(t+1)}, \bm{u}_n^{(t+1)}) = (\bm{\theta}_n^{(t)}, \bm{u}_n^{(t)})$
\EndIf
\end{algorithmic}
\end{algorithm}

\subsection{\texorpdfstring{Proposal distributions $q_u$}{Proposal distribution for U}} \label{sec:proposal_u}

This section introduces the proposal distribution $q_u$ adopted for $\bu_n$, adaptive to the noise affecting the observation $(n, \ell) \in \llbracket1, N\rrbracket \times \llbracket1, L\rrbracket$. 

Given a candidate $\tilde{\btheta}_n^{(t+1)}$, the candidate $\tilde{\bu}_n^{(t+1)}$ needs to be proposed such that each component $\tilde{u}\nl^{(t+1)}$ lies in a high probability region for the conditional distribution of $u\nl$, whose pdf is of the form
\begin{equation}
  \pi(u\nl \mid y\nl, \tilde{\btheta}_n^{(t+1)}) \propto \pi(y\nl \mid u\nl) \: \pi(u\nl \mid \tilde{\btheta}_n^{(t+1)}).
  \label{eq:cond_u}
\end{equation}
A proposal distribution should then be chosen to well approximate \eqref{eq:cond_u}, which is not a standard distribution.
\Cref{eq:cond_u} further suggests that the proposal distribution $q_u$ could leverage information related to both $y_{n,\ell}$ and $\tilde{\boldsymbol{\theta}}_n^{(t+1)}$.
In the case of censored data, i.e. $y_{n,\ell} = \omega_{n,\ell}$, the observation $y_{n,\ell}$ is poorly informative. Two different proposal distributions are used depending on whether the observation is censored or not.

\subsubsection{Non-censored observations} 
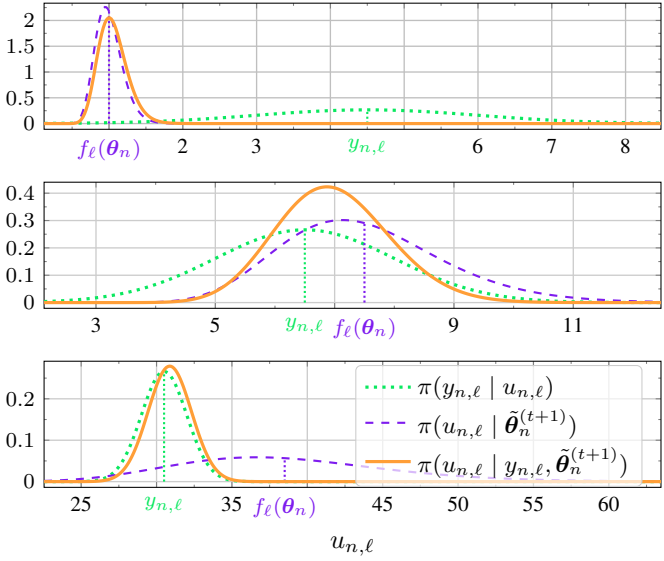
\begin{figure}
\centering
\begin{tikzpicture}
\begin{groupplot}[
    group style={
        group size=1 by 3,
        vertical sep=.7cm
    },
    width=1.1\linewidth, height=3.25cm,
    samples=200,
    legend style={
        font=\small,
        rounded corners=2pt,
        fill=white,
        fill opacity=0.7,
        text opacity=1,
        draw={gray!40},
        align=left,
        legend cell align=left,
    },
    legend pos=north east,
    tick style={major tick length=2pt, minor tick length=1pt, thin},
    tick label style={font=\footnotesize},
    minor tick num=1,
    extra x tick style={
        grid=none, 
        tick style={draw=none},
        tick label style={font=\footnotesize\bfseries, yshift=0pt}
    }
]

% High regime
\nextgroupplot[
    title style={yshift=-1.ex},
    xmin=0.12, xmax=8.48,
    domain=0.01:8.48,
    samples=400,
    ymin=-0.111, ymax=2.335,
    xtick={0,1,2,3,4,5,6,7,8},
    xticklabels={0,\empty,2,3,\empty,\empty,6,7,8}, 
    extra x ticks={1.0, 4.5},
    extra x tick labels={\textcolor{mypurple}{$f_\ell(\boldsymbol{\theta}_n)$}, \textcolor{mycyan}{$y_{n,\ell}$}},
    ytick={-0.5,0.0,0.5,1.0,1.5,2.0,2.5},
    grid=major,
    major grid style={line width=0.5pt,draw=gray!50},
]

% Additive noise PDF
\addplot[mycyan,very thick,dotted,forget plot] {1/(sqrt(2*pi*1.5^2)) * exp(-0.5*((x-4.5)/1.5)^2)};
\draw[mycyan, thick, densely dotted] (axis cs:4.5,0) -- (axis cs:4.5,0.266);

% Multiplicative noise PDF
\addplot[mypurple,thick,dashed,forget plot] {1/(x*0.18232*sqrt(2*pi)) * exp(- (ln(x) - (-0.01662))^2 / (2*0.18232^2))};
\draw[mypurple, thick, densely dotted] (axis cs:1.0,0) -- (axis cs:1.0,2.188);

% Conditional PDF
\addplot[myorange,very thick,forget plot]
  {((1/(sqrt(2*pi*1.5^2)) * exp(-0.5*((x-4.5)/1.5)^2)) *
    (1/(x*0.18232*sqrt(2*pi)) * exp(- (ln(x) - (-0.01662))^2 / (2*0.18232^2)))) / 0.018570239509210624};

% Medium regime
\nextgroupplot[
    title style={yshift=-1.ex},
    xmin=2.13, xmax=12.47,
    domain=2.13:12.47,
    samples=400,
    ymin=-0.02096, ymax=0.44012,
    xtick={3,5,7,9,11}, 
    xticklabels={3,5,\empty,9,11},
    extra x ticks={6.5, 7.5},
    extra x tick labels={\textcolor{mycyan}{$y_{n,\ell}$}, \textcolor{mypurple}{$f_\ell(\boldsymbol{\theta}_n)$}},
    ytick={-0.1,0,0.1,0.2,0.3,0.4,0.5},
    grid=major,
    major grid style={line width=0.5pt,draw=gray!40},
]

% Additive noise PDF
\addplot[mycyan,very thick,dotted,forget plot] {1/(sqrt(2*pi*1.5^2)) * exp(-0.5*((x-6.5)/1.5)^2)};
\draw[mycyan, thick, densely dotted] (axis cs:6.5,0) -- (axis cs:6.5,0.266);

% Multiplicative noise PDF
\addplot[mypurple,thick,dashed,forget plot] {1/(x*0.18232*sqrt(2*pi)) * exp(- (ln(x) - 1.99828)^2 / (2*0.18232^2))};
\draw[mypurple, thick, densely dotted] (axis cs:7.5,0) -- (axis cs:7.5,0.292);

% Conditional PDF
\addplot[myorange,very thick]
  {((1/(sqrt(2*pi*1.5^2)) * exp(-0.5*((x-6.5)/1.5)^2)) *
    (1/(x*0.18232*sqrt(2*pi)) * exp(- (ln(x) - 1.99828)^2 / (2*0.18232^2)))) / 0.18};

% Low regime
\nextgroupplot[
    title style={yshift=-1.ex},
    xmin=22.54, xmax=63.46,
    domain=22.54:63.46,
    samples=400,
    ymin=-0.01385, ymax=0.29078,
    xtick={20,25,30,35,40,45,50,55,60,65},
    xticklabels={20,25,\empty,35,\empty,45,50,55,60,65},
    extra x ticks={30.5, 38.5},
    extra x tick labels={\textcolor{mycyan}{$y_{n,\ell}$}, \textcolor{mypurple}{$f_\ell(\boldsymbol{\theta}_n)$}},
    ytick={-0.1,0,0.1,0.2,0.3},
    xlabel={$u\nl$},
    grid=major,
    major grid style={line width=0.5pt,draw=gray!40},
]

% Additive noise PDF
\addlegendimage{mycyan,very thick,dotted}
\addlegendentry{$\pi(y_{n,\ell}\mid u_{n,\ell})$}
\addplot[mycyan,very thick,dotted,forget plot] {1/(sqrt(2*pi*1.5^2)) * exp(-0.5*((x-30.5)/1.5)^2)};
\draw[mycyan, thick, densely dotted] (axis cs:30.5,0) -- (axis cs:30.5,0.266);

% Multiplicative noise PDF
\addlegendimage{mypurple,thick,dashed}
\addlegendentry{$\pi(u_{n,\ell} \mid \tilde{\boldsymbol{\theta}}_n^{(t+1)})$}
\addplot[mypurple,thick,dashed,forget plot] {1/(x*0.18232*sqrt(2*pi)) * exp(- (ln(x) - 3.63404)^2 / (2*0.18232^2))};
\draw[mypurple, thick, densely dotted] (axis cs:38.5,0) -- (axis cs:38.5,0.057);

% Conditional PDF
\addlegendimage{myorange,very thick}
\addlegendentry{$\pi(u_{n,\ell} \mid y_{n,\ell}, \tilde{\boldsymbol{\theta}}_n^{(t+1)})$}
\addplot[myorange,very thick]
    {((1/(sqrt(2*pi*1.5^2)) * exp(-0.5*((x-30.5)/1.5)^2)) *
        (1/(x*0.1823215567939546*sqrt(2*pi)) * exp(- (ln(x) - 3.634037666257853)^2 / (2*0.1823215567939546^2)))) / 0.035};

\end{groupplot}
\end{tikzpicture}
    \caption{Impact of the amplitude of $f_l(\btheta_n)$ on the dominant noise regime in the pdf of the conditional distribution on $u\nl$. \textit{Top:} $\pi(u\nl \mid \tilde{\btheta}_n^{(t+1)})$ dominates. \textit{Middle:} both component distributions are significant. \textit{Bottom:} $\pi(y\nl\mid u\nl)$ dominates.}
    \label{fig:proposal_U}
\end{figure}

For non-censored observations, both the observation $y\nl$ and $f_{\ell}(\tilde{\btheta}_n^{(t+1)})$ are informative and can be used to guide the proposal distribution. 
\Cref{fig:proposal_U} illustrates the typical situations. \Cref{fig:proposal_U}(top \& bottom) show cases where \eqref{eq:cond_u} is guided by the distribution factor with smaller variance, but with different tails behaviors. 
\Cref{fig:proposal_U}(middle) illustrates an example where both distributions have similar variances around close $y\nl$ and $f_{\ell}(\tilde{\btheta}_n)$. Then the conditional distribution depends on both  component distributions in a balanced manner.  Many other scenarios are difficult to characterize a priori.

This difficulty is tackled by using a parametric distribution whose parameters are fitted to match the conditional distribution~\eqref{eq:cond_u}.
Note that the distribution \eqref{eq:cond_u} is restricted to positive values, and is skewed due to the lognormal term $\pi(u\nl \mid \tilde{\btheta}_n^{(t+1)})$. 
The larger the scale parameter $\sigma_m$ of the lognormal distribution, the more pronounced the skewness. 
However, even for moderate values of $\sigma_m$, the conditional \eqref{eq:cond_u} can still be highly skewed when $f_\ell(\tilde{\btheta}_n^{(t+1)})$ significantly differs from $y\nl$. 
This situation may occur since the I-MTM proposal $q_{\btheta}^{(2)}$ allows jumps to distant locations in the parameter space, possibly generating a candidate $f_\ell(\tilde{\btheta}_n^{(t+1)})$ in the tails of $\pi(y\nl\mid u\nl)$. 
It is therefore crucial to consider a skewed proposal distribution $q_u(u\nl \mid \tilde{\btheta}_n^{(t+1)}, y\nl)$. 
In practice, the dimension $L$ can be much larger than $D$ so that  
the proposal distribution should also be cheap and easy to sample from.% in order to limit the computational cost.

We adopt a Gamma proposal for $u\nl$ in the non-censored case to better capture the tail behavior of (\ref{eq:cond_u}). Specifically, the Gaussian likelihood $\pi(y\nl \mid u\nl)$ modifies the lognormal prior $\pi(u\nl \mid \tilde{\btheta}_n^{(t+1)})$ in two ways: it accelerates the decay in the right tail, and mitigates the rapid decrease near zero in the left tail.
For a given mean and variance, the Gamma distribution naturally exhibits a lighter right tail and a heavier left tail compared to the lognormal distribution, making it a good approximation.
For completeness, details on the procedure adopted to fit the parameters of the Gamma proposal to~\eqref{eq:cond_u} at a limited cost are provided in Appendix~\ref{sec:fitting_procedure_proposal_u}.
% This proposal distribution has been added to the \texttt{BEETROOTS} library\footnote[1]{The code is available at \url{https://github.com/pierrePalud/beetroots}} and can be used as a black box method.

\subsubsection{Censored observations} For censored observations, the term $\pi(y\nl \mid u\nl)$ in \eqref{eq:cond_u} is
\begin{equation}
  \pi(y\nl \mid u\nl) =  \Phi(\omega\nl \mid u\nl, \, \sigma_{a,n,\ell}^2) \delta(y\nl - \omega\nl).
\end{equation}
The observation $y\nl$ is therefore poorly informative about $u\nl$. The conditional distribution \eqref{eq:cond_u} is mostly guided by the term $\pi(u\nl \mid \tilde{\btheta}_n^{(t+1)})$. In this case, ancestral sampling \cite{Bishop_2006} appears as a natural choice to propose $u\nl$, i.e.
\begin{equation}
    q_u(u\nl \mid \tilde{\btheta}_n^{(t+1)}) = \pi(u\nl \mid \tilde{\btheta}_n^{(t+1)})
\end{equation}
so that
\begin{equation}
    \tilde{u}\nl^{(t+1)} \mid \tilde{\btheta}_n^{(t+1)} \sim \log\mathcal{N}\bigl(\log\bigl(f_{\ell}(\tilde{\btheta}_n^{(t+1)})\bigr) - \sigma_m^2/2, \sigma_m^2\bigr).
\end{equation}

% \ngoe{
% \section{Modelling decision tool}
% Intro
% \begin{itemize}
%     \item Need for complex models depends on:
%     \begin{itemize}
%         \item Range captured by the forward model
%         \item Noise levels
%     \end{itemize}
%     \item Need for a decision tool to guide the choice of model complexity
% \end{itemize}
% Method
% \begin{itemize}
%     \item Introduce motivations for KS-based statistics (multiple decades of amplitude)
%     \item Definition of KS statistic
%     \item Definition of KS-based metrics for model selection (expectation)
% \end{itemize}
% }

\section{Numerical experiments}
\label{sec:experiments}

The proposed hierarchical approach has been numerically evaluated on the challenging astrophysical inverse problem considered in~\cite{Palud_Bron_Chainais_Petit_Thouvenin_Santa-Maria_Goicoechea_Languignon_Gerin_Pety_et_al_2025}.
This problem combines all the challenges mentioned in \Cref{sec:intro}, i.e. multiplicative and additive sources of noise, censoring, multimodality induced by the non-linearity of the forward operator, and high dynamic ranges for both the parameters and the observations. The proposed approach is compared with the approximate approach from~\cite{Palud_Thouvenin_Chainais_Bron_LePetit_2023}, referred to as the \emph{interpolated likelihood}. Comparisons are based on a revised implementation of \cite{Palud_Thouvenin_Chainais_Bron_LePetit_2023}, fixing an issue related to the normalizing constant on $\bTheta$ for the interpolated likelihood function.
Further details on this issue and the solution proposed are provided in the supplementary material. We also compare the proposed approach with purely additive and purely multiplicative likelihood models which are referred to as \emph{additive} and \emph{multiplicative} likelihoods, respectively.
All the experiments have been run on a computer equipped with an Intel(R) Xeon(R) w9-3475X CPU, with 251 GB of RAM. Codes to reproduce the experiments are part of a forthcoming update of the \texttt{BEETROOTS} library\footnote[1]{Code available at \url{https://github.com/pierrePalud/beetroots/tree/hierarchical}}~\cite{Palud_Bron_Chainais_Petit_Thouvenin_Santa-Maria_Goicoechea_Languignon_Gerin_Pety_et_al_2025}.

\subsection{Experiment setting}
\begin{figure}
  \centering
  
  \begin{subfigure}[b]{0.26\linewidth}
    \centering
    \includegraphics[height=2.2cm]{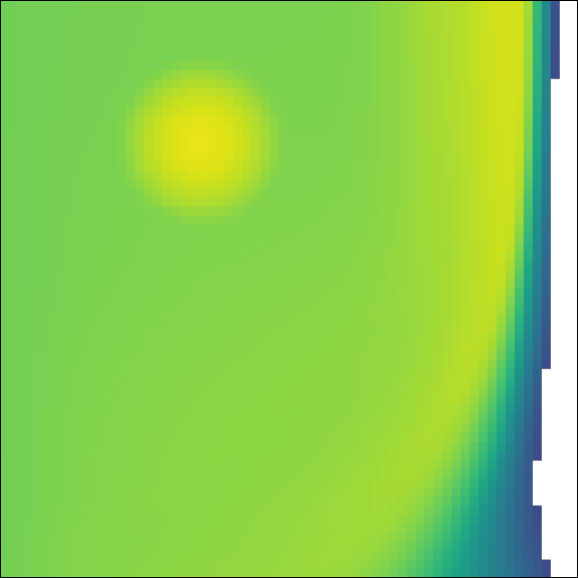}
    \caption{$\ell=1$}
    \label{fig:synth_l0}
  \end{subfigure}
  \hfill
  \begin{subfigure}[b]{0.35\linewidth}
    \centering
    \includegraphics[height=2.2cm]{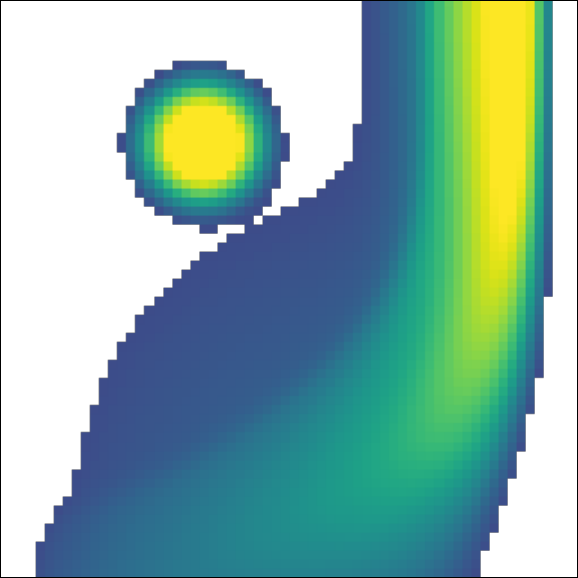}
    % \hspace{0.05cm}
    \includegraphics[height=2.2cm]{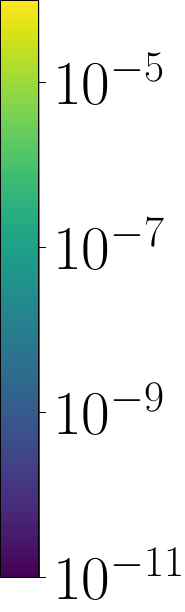}
    \caption{$\ell=8$}
    \label{fig:synth_l7}
  \end{subfigure}
  \hfill
  \begin{subfigure}[b]{0.35\linewidth}
    \centering
    \includegraphics[height=2.2cm]{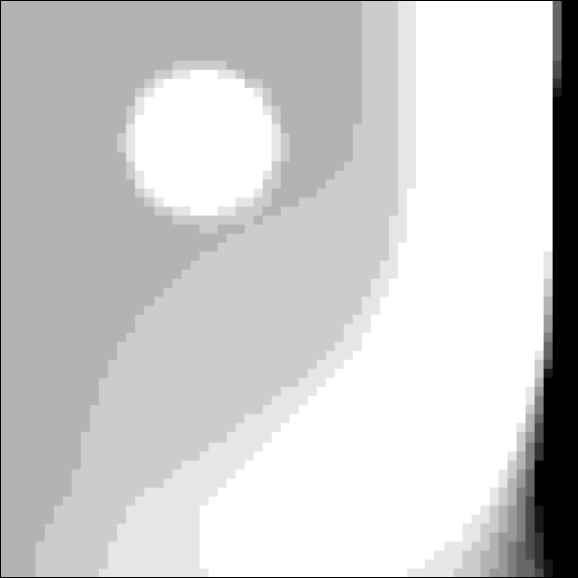}
    % \hspace{0.05cm}
    \includegraphics[height=2.2cm]{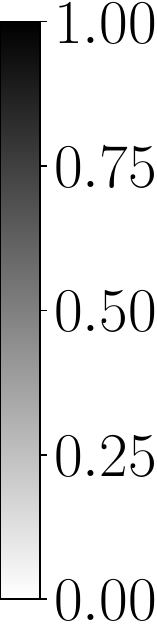}
    \caption{Censoring map}
    \label{fig:synth_censoring}
  \end{subfigure}
  
  \caption{Examples of synthetic observations for (a) $\ell=1$ and (b) $\ell=8$, along with (c) the rate of censored lines per pixel. In the observation maps, white pixels represent censored observations fixed to $y\nl = \omega\nl$.}
  \label{fig:synthetic_data}
\end{figure}

Synthetic observation maps of an interstellar cloud $\bY \in \mathbb{R}^{N\times L}$ of $N$ pixels and $L$ spectral bands are generated from realistic maps of parameters of interest $\bTheta^* \in \mathbb{R}^{N\times D}$ using \eqref{eq:obs-model}. The ground truth parameters $\bTheta^*$ are linked to physical parameters $\boldsymbol{\Phi}^* = (\bm{\phi}^*)_{n=1}^N$\footnote[2]{Here, $\boldsymbol{\Phi}^* = (\boldsymbol{\kappa}^*, \boldsymbol{P}_{th}^*, \boldsymbol{G}_0^*, \boldsymbol{A}_V^*) \in \mathbb{R}^{N\times D}$ gather respectively a nuisance parameter linked to the acquisition procedure, the thermal pressure, the intensity of a UV radiative field and the interstellar cloud depth along the telescope line of sight} through \eqref{eq:normalized_space}. The parameters are estimated in this log-normalized space to handle the high dynamic range of the physical parameters $\boldsymbol{\Phi}^*$. The true forward operator $\boldsymbol{f}^*$ is a highly non-linear and costly numerical model \cite{Petit_Nehme_Bourlot_Roueff_2006} replaced by a surrogate neural network model $\bm{f}$ \cite{Palud_Einig_Le_Petit_Bron_Chainais_Chanussot_Pety_Thouvenin_Languignon_Beslic_et_al_2023}, faster to evaluate and differentiable to accommodate the PMALA proposal.
In these experiments, the dimensions are set to $N=64$, $L=10$ and $D=4$.
The standard deviation of the additive white Gaussian noise is fixed to $\sigma_{a, n, \ell} = 1.39\times 10^{-10}, \ \forall (n,\ell) \in \llbracket 1, N \rrbracket \times \llbracket 1, L \rrbracket$, which is a typical and realistic value for instrumental noise. The scale parameter of the lognormal multiplicative noise is set to 3 different values $\exp(\sigma_m) \in \{1.1, 1.5, 2.0\}$, representing a low ($\approx 10\%$), a moderate ($\approx 40\%$) and a high ($\approx 80\%$) level of multiplicative noise. 
Finally, a censoring threshold $\omega_{n,\ell} = 3\sigma_{a,n,\ell}$ is applied to the observations to account for detection limits.
Censored observations represent around 20\% of the $N\times L$ observations, and are located in specific regions of the maps.
\Cref{fig:synthetic_data} shows examples of synthetic observations and the pixel-wise proportion of censored observations.
The spatial regularization hyperparameters are set to $\tau_d = 20$ for all $d \in \llbracket 1, D \rrbracket$. 
The value has been tuned manually to provide a good trade-off between smoothness and detail preservation. 
The parameter $\delta$ in \eqref{eq:smooth_indicator} is set to $\delta = 10^{4}$. For any $d \in \llbracket 1, D \rrbracket$, the validity intervals $\mathcal{C}_d$ on the normalized parameters $\bTheta$  are defined using \eqref{eq:normalized_space}, with the validity intervals on the physical parameters $\boldsymbol{\Phi}$ given in \cite{Palud_Bron_Chainais_Petit_Thouvenin_Santa-Maria_Goicoechea_Languignon_Gerin_Pety_et_al_2025}. 
% and the log-normalization process \eqref{eq:normalized_space}.

\subsection{Evaluation metrics}

Performance is evaluated in terms of reconstruction quality, uncertainty quantification, and computational cost. The latter is evaluated in terms of computing time per iteration.

\textbf{Reconstruction metric --}
The reconstruction quality is assessed using the mean absolute error (MAE) defined, for each estimated parameter map 
%$\widehat{\bTheta}_{\cdot, d} = (\widehat{\theta}_{n,d})_{n=1}^N$, 
as
\begin{equation}
  \mathrm{MAE}(\widehat{\bTheta}_{\cdot, d}, \bTheta^*_{\cdot, d}) =  \frac{1}{N}\sum_{n=1}^{N}\lvert  \widehat{\theta}\nd - \theta\nd^* \rvert,
\end{equation}
where $\widehat{\bTheta}$ is an estimate of $\bTheta^*$. This metric is defined in the log-normalized space, but can be interpreted in terms of the physical parameters by looking at the first order Taylor expansion of \eqref{eq:normalized_space}
\begin{equation}
  \label{eq:taylor_expansion_mae}
  \nu(\widehat{\phi}\nd) = \nu(\phi\nd^*) + \frac{1}{\ln(10)\sigma_d} \frac{\widehat{\phi}\nd - \phi\nd^*}{\phi\nd^*} + \mathcal{O}(\lvert \widehat{\phi}\nd - \phi\nd^* \rvert^2).
\end{equation}
The MAE with respect to the log-normalized parameters $\bTheta$ is therefore closely related to a relative error with respect to $\boldsymbol{\Phi}$, up to a multiplicative constant
\begin{equation}
  \lvert \widehat{\theta}\nd - \theta\nd^* \rvert \approx \frac{1}{\ln(10)\sigma_d} \bigg\lvert\frac{\widehat{\phi}\nd - \phi\nd^* }{\phi\nd^*}\bigg\rvert.
\end{equation}

The minimum mean square error (MMSE) estimator is used for $\widehat{\bTheta}$, and is defined as
\begin{equation}
  \widehat{\bTheta} = \frac{1}{T-T_{BI}}\sum_{t=T_{BI}+1}^{T} \bTheta^{(t)},
\end{equation}
where $T$ and $T_{BI}$ denote the total number of iterations and the burn-in period, respectively. 

\textbf{Uncertainty quantification metric --}
Uncertainty is quantified \emph{via} entry-wise 95\% credible interval (CI) of the samples $\{\bTheta^{(t)}\}_{t=T_{BI}+1}^{T}$. Note that CI is equivariant through the log-normalization \eqref{eq:normalized_space}, meaning that a CI of level $\beta$ for $\bTheta$ can be transformed into a CI of level $\beta$ for $\boldsymbol{\Phi}$ by inverting \eqref{eq:normalized_space}.

\subsection{Model selection}

A model selection analysis is performed to compare the proposed hierarchical model with the three other models. Model selection is crucial in absence of ground truth data to distinguish between different models, taking into account their associated uncertainties.
Model selection methods can be categorized into two main families: Bayes factors based methods~\cite{Kass_Raftery_1995} and predictive performance based methods~\cite{Gelman_Hwang_Vehtari_2014}. The former are often considered as the gold standard, but appears computationally challenging and prone to errors in practice, especially for high dimensional problems. The latter are aimed at estimating the expected log predictive density (ELPD) of the model, defined by
\begin{align}
    \textrm{ELPD}\nl &= \mathbb{E}_{\pi(\tilde{y}\nl \mid \btheta_n^*)}\bigl[ \log \pi(\tilde{y}\nl \vert y\nl)\bigr], \\
    \textrm{ELPD} &= \sum_{n=1}^{N} \sum_{\ell=1}^{L} \textrm{ELPD}\nl,
\end{align}
where $\pi(\tilde{y}\nl \mid \btheta_n^*)$ represents the true generative distribution of the data \eqref{eq:obs-model}. The higher the ELPD, the better the predictive performance of the model considered. Maximizing the ELPD amounts to minimizing the Kullback-Leibler divergence between the true generative distribution of the data and the predictive distribution of the model \cite{Gelman_Hwang_Vehtari_2014}. The ELPD metric evaluates each model by looking at the distribution as a whole, instead of focusing on a point estimate as the reconstruction metrics.
The use of synthetic data enables to estimate the ELPD efficiently by Monte Carlo methods. It boils down to a 1D integral estimation in the considered experiment, thanks to the factorization of the likelihood distribution. Details on the estimation of the ELPD for all likelihood models are provided in the supplementary material.
The different likelihoods are compared by computing the pairwise difference of ELPD maps as follows
\begin{equation}
    \Delta\textrm{ELPD}\nl^{(i,j)} = \widehat{\textrm{ELPD}}^{(i)}\nl - \widehat{\textrm{ELPD}}^{(j)}\nl, \quad i \ne j,
\end{equation}
where $\widehat{\textrm{ELPD}}^{(i)}\nl$ and $\widehat{\textrm{ELPD}}^{(j)}\nl$ are the $(n,\ell)$-th entries of the estimated ELPD maps for the models $i$ and $j$. A positive value for $\Delta\textrm{ELPD}\nl^{(i,j)}$ means that model $i$ has a better predictive performance than $j$ for the $(n,\ell)$-th observation.

\subsection{Sampling parameters}

The total number of iterations is set to $T=10\,000$ for all approaches, including a burn-in period of $T_{BI} = 1\,500$.
Both the proposed hierarchical approach and \cite{Palud_Thouvenin_Chainais_Bron_LePetit_2023} have been compared with the same algorithm parameters detailed below. 
The probability of using the MH kernel over the I-MTM kernel is set to $p=0.5$. 
The step size in the PMALA proposal is set to $\epsilon = 10^{-2}$, the damping parameter to $\eta = 10^{-5}$ and the RMSProp decay rate to $\alpha = 0.5$. 
The number of candidates in I-MTM is set to $M=50$. 
% These parameters are the same for both all approaches. 

\section{Results and discussion}
This section presents the results of the numerical experiments run with the setup described in \Cref{sec:experiments}. All algorithms have been run 5 times with different random initialization seeds to assess the robustness of the results. All the seeds led to consistent results. The estimates obtained for a single seed with $\sigma_m=\log(1.5)$ are displayed in the main paper for the sake of brevity. The estimates for all multiplicative noise levels $\sigma_m$ and the same seed are displayed in the supplementary material.
\Cref{fig:mmse_reconstruction_maps} shows the MMSE estimates of each parameter map $\bTheta_{\cdot, d}$ for all likelihood models. 
\Cref{tab:mae_comparison} summarizes the averaged MAE metric over the 5 runs for each of the four models and for different noise levels $\sigma_m$.
All the approaches show similar performances in terms of reconstruction quality for a low multiplicative noise level $\sigma_m$. As $\sigma_m$ increases, the additive approach tends to perform worse than the others. The supplementary material provides results for other noise and censoring levels. The relative performance of the additive and multiplicative models is sensitive to these scenarios. The better choice between these two models is difficult to know a priori since the dominant source noise might vary across the $L$ spectral lines.

In contrast, both the interpolated and hierarchical models provide similar good reconstruction performance in all scenarios. We emphasize that only the proposed exact hierarchical model can be readily applied with reliable performance to all of these scenarios without additional hyperparameters tuning, making it a jackknife model.

\begin{figure}
\centering
\includegraphics[width=1\linewidth]{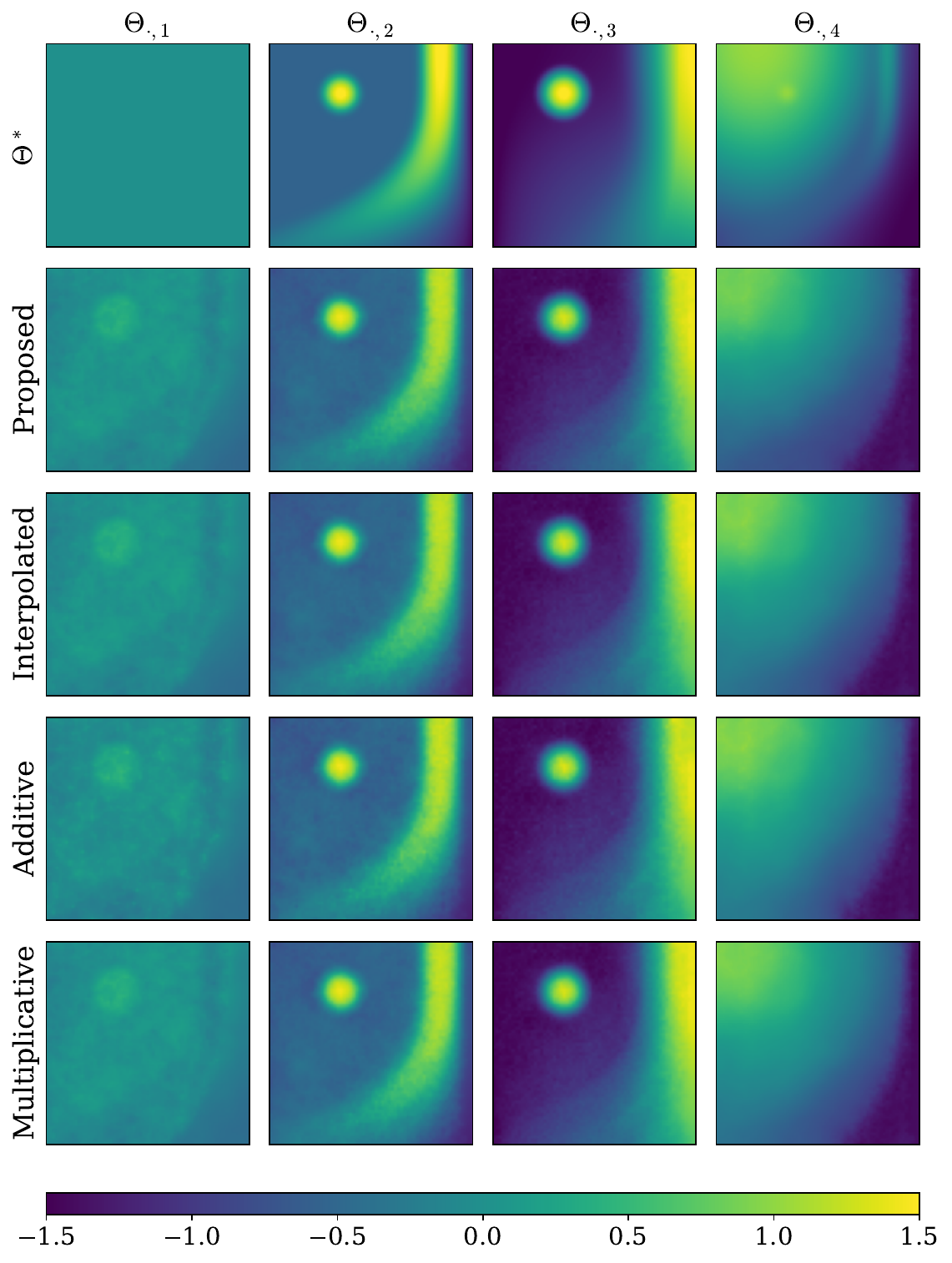}
\caption{
Reconstruction results (MMSE) for each parameter map and for all likelihood models with $\sigma_m = \log(1.5)$. From top to bottom: ground truth, hierarchical, interpolated, additive, multiplicative.}
\label{fig:mmse_reconstruction_maps}
\end{figure}

\begin{table}[t]
\centering
\caption{MAE (mean $\pm$ s.t.d.) comparison across the different models and noise levels $\sigma_m$. Results are scaled by a factor of $100$ for readability.}
\label{tab:mae_comparison}
\renewcommand{\arraystretch}{1.2} 

\begin{tabular}{clcccc}
\toprule
 & & \multicolumn{4}{c}{\textbf{MAE} $\downarrow$} \\
\cmidrule(lr){3-6}
 $\sigma_m$ & \textbf{Model} & $\bTheta_{\cdot, 0}$ & $\bTheta_{\cdot, 1}$ & $\bTheta_{\cdot, 2}$ & $\bTheta_{\cdot, 3}$ \\ \midrule
\multirow{5}{*}{$\log(1.1)$} & Additive        & \makecell{$5.74$ \\ [-0.8ex] {\scriptsize $\pm 0.11$}} & \makecell{$5.09$ \\ [-0.8ex] {\scriptsize $\pm 0.02$}} & \makecell{$4.37$ \\ [-0.8ex] {\scriptsize $\pm 0.03$}} & \makecell{$10.72$ \\ [-0.8ex] {\scriptsize $\pm 0.13$}} \\
 & Multiplicative  & \makecell{$5.70$ \\ [-0.8ex] {\scriptsize $\pm 0.17$}} & \makecell{$4.99$ \\ [-0.8ex] {\scriptsize $\pm 0.01$}} & \makecell{$4.45$ \\ [-0.8ex] {\scriptsize $\pm 0.03$}} & \makecell{$11.14$ \\ [-0.8ex] {\scriptsize $\pm 0.22$}} \\
 & Interpolation   & \makecell{$5.83$ \\ [-0.8ex] {\scriptsize $\pm 0.21$}} & \makecell{$5.06$ \\ [-0.8ex] {\scriptsize $\pm 0.01$}} & \makecell{$4.38$ \\ [-0.8ex] {\scriptsize $\pm 0.04$}} & \makecell{$10.86$ \\ [-0.8ex] {\scriptsize $\pm 0.19$}} \\
 & Hierarchical    & \makecell{$5.71$ \\ [-0.8ex] {\scriptsize $\pm 0.21$}} & \makecell{$5.03$ \\ [-0.8ex] {\scriptsize $\pm 0.02$}} & \makecell{$4.43$ \\ [-0.8ex] {\scriptsize $\pm 0.02$}} & \makecell{$10.99$ \\ [-0.8ex] {\scriptsize $\pm 0.13$}} \\
 \hline
\multirow{5}{*}{$\log(1.5)$} & Additive        & \makecell{$12.33$ \\ [-0.8ex] {\scriptsize $\pm 0.37$}} & \makecell{$8.59$ \\ [-0.8ex] {\scriptsize $\pm 0.06$}} & \makecell{$7.64$ \\ [-0.8ex] {\scriptsize $\pm 0.09$}} & \makecell{$13.41$ \\ [-0.8ex] {\scriptsize $\pm 0.26$}} \\
 & Multiplicative  & \makecell{$11.07$ \\ [-0.8ex] {\scriptsize $\pm 0.30$}} & \makecell{$8.44$ \\ [-0.8ex] {\scriptsize $\pm 0.07$}} & \makecell{$7.59$ \\ [-0.8ex] {\scriptsize $\pm 0.06$}} & \makecell{$14.22$ \\ [-0.8ex] {\scriptsize $\pm 0.53$}} \\
 & Interpolation   & \makecell{$11.53$ \\ [-0.8ex] {\scriptsize $\pm 0.25$}} & \makecell{$8.75$ \\ [-0.8ex] {\scriptsize $\pm 0.07$}} & \makecell{$7.75$ \\ [-0.8ex] {\scriptsize $\pm 0.04$}} & \makecell{$12.76$ \\ [-0.8ex] {\scriptsize $\pm 1.14$}} \\
 & Hierarchical    & \makecell{$11.55$ \\ [-0.8ex] {\scriptsize $\pm 0.23$}} & \makecell{$8.61$ \\ [-0.8ex] {\scriptsize $\pm 0.07$}} & \makecell{$7.69$ \\ [-0.8ex] {\scriptsize $\pm 0.04$}} & \makecell{$13.32$ \\ [-0.8ex] {\scriptsize $\pm 0.64$}} \\
 \hline
\multirow{5}{*}{$\log(2.0)$} & Additive        & \makecell{$16.90$ \\ [-0.8ex] {\scriptsize $\pm 0.51$}} & \makecell{$11.15$ \\ [-0.8ex] {\scriptsize $\pm 0.06$}} & \makecell{$9.94$ \\ [-0.8ex] {\scriptsize $\pm 0.13$}} & \makecell{$15.24$ \\ [-0.8ex] {\scriptsize $\pm 0.63$}} \\
 & Multiplicative  & \makecell{$13.05$ \\ [-0.8ex] {\scriptsize $\pm 0.26$}} & \makecell{$10.65$ \\ [-0.8ex] {\scriptsize $\pm 0.10$}} & \makecell{$9.31$ \\ [-0.8ex] {\scriptsize $\pm 0.10$}} & \makecell{$16.09$ \\ [-0.8ex] {\scriptsize $\pm 0.83$}} \\
 & Interpolation   & \makecell{$13.63$ \\ [-0.8ex] {\scriptsize $\pm 0.41$}} & \makecell{$11.02$ \\ [-0.8ex] {\scriptsize $\pm 0.13$}} & \makecell{$9.53$ \\ [-0.8ex] {\scriptsize $\pm 0.08$}} & \makecell{$15.05$ \\ [-0.8ex] {\scriptsize $\pm 0.72$}} \\
 & Hierarchical    & \makecell{$14.29$ \\ [-0.8ex] {\scriptsize $\pm 0.48$}} & \makecell{$10.72$ \\ [-0.8ex] {\scriptsize $\pm 0.13$}} & \makecell{$9.42$ \\ [-0.8ex] {\scriptsize $\pm 0.06$}} & \makecell{$15.92$ \\ [-0.8ex] {\scriptsize $\pm 0.66$}} \\
\bottomrule
\end{tabular}
\end{table}

\begin{figure}
\centering
\includegraphics[width=1\linewidth]{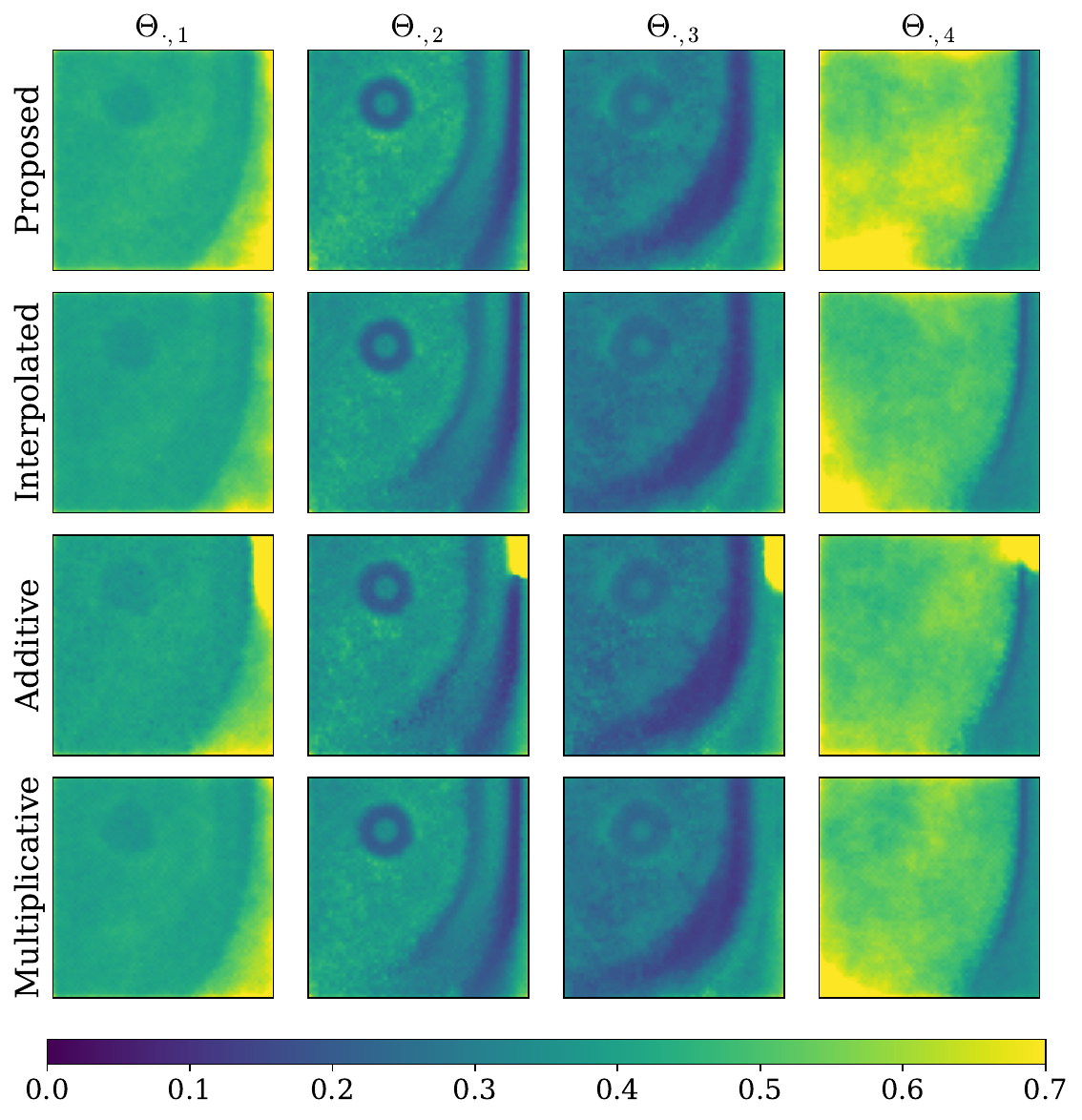}
\caption{Element wise CI size for each parameter map and for all likelihood models and for $\sigma_m = \log(1.5)$. From top to bottom: ground truth, hierarchical, interpolated, additive, multiplicative.}
\label{fig:ci_factor_maps}
\end{figure}

\Cref{fig:ci_factor_maps} displays the size of the corresponding elementwise CI maps for the 4 physical parameters for all four models.
Similar global behaviors are observed for all approaches. The additive model shows a clear failure in the top right corner on 4 seeds out of the 5 used.
This region corresponds to a high censoring rate area (see \Cref{fig:synthetic_data}), where the observations are absent or weakly informative. 
It also corresponds to an area of sharp transition in the amplitude of the ground truth parameters. 
This behavior is only moderately compatible with the spatial regularization considered in this work. 
The three other models are able to accurately model the high SNR region and do not exhibit such issue.

\begin{figure}
  \includegraphics[width=\linewidth]{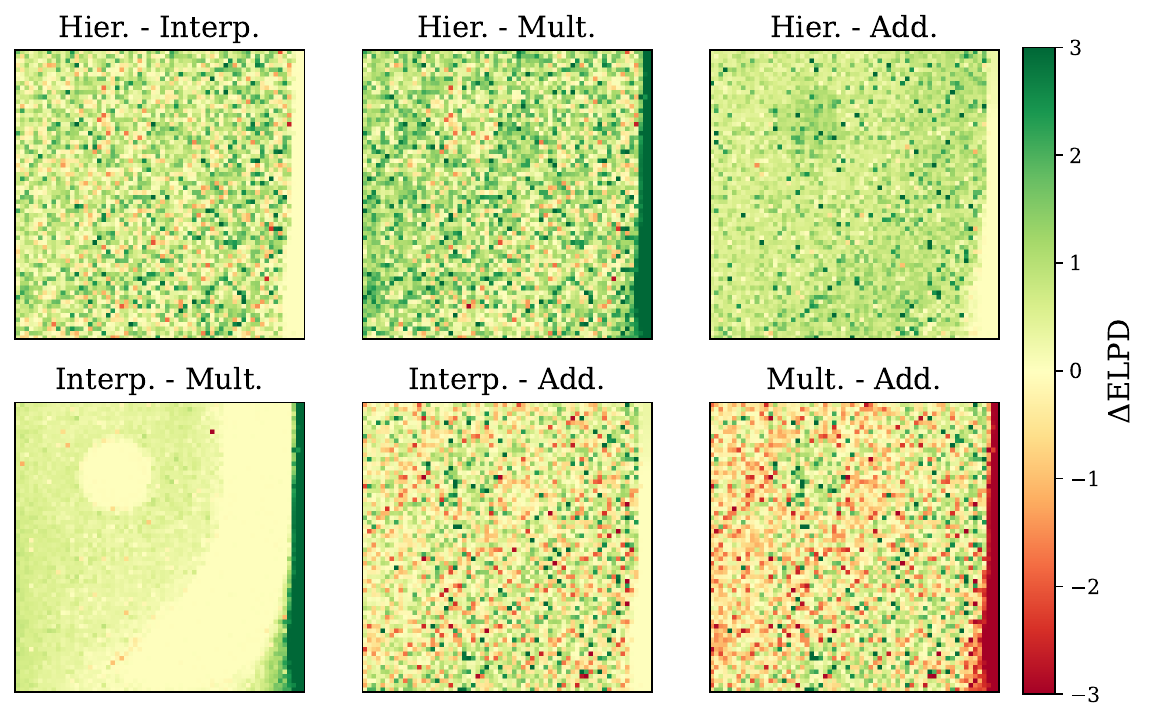}
  \caption{$\Delta\textrm{ELPD}$ maps for $\sigma_m=\log(1.5)$ and for each pair of likelihood models. The $L$ lines are summed for visual purposes. First row: hierarchical vs interpolation/multiplicative/additive (from left to right). Second row: interpolation vs multiplicative/additive, multiplicative vs additive (from left to right). Positive values (in green) indicate a better performance of the first model in the pair, while lower values (in red) indicate a better performance of the second model.}
  \label{fig:elpd_pairwise_diff_sigma_1p5}
\end{figure}

\Cref{fig:elpd_pairwise_diff_sigma_1p5} shows the $\Delta\textrm{ELPD}$ maps for each pair of likelihood models under $\sigma_m=\log(1.5)$. \Cref{tab:elpd_pairwise_noise} summarizes these maps by providing the mean of all $\Delta\textrm{ELPD}\nl^{(i,j)}$ values, \emph{i.e.} $\textrm{ELPD}/NL$, for each pair of models $(i,j)$ and for all three multiplicative noise levels $\sigma_m$. The results in \Cref{tab:elpd_pairwise_noise} give an indicative quantification of the relative predictive performance of each model. For instance, the first row of \Cref{tab:elpd_pairwise_noise} shows that the hierarchical model is $\exp(0.182) \approx 1.2$ times more likely on average than the interpolated model %at the elementwise level 
for $\sigma_m=\log(2.0)$.
These results highlight the superior predictive performance of the hierarchical model over the other models, especially at higher noise levels. \Cref{fig:elpd_pairwise_diff_sigma_1p5} also highlights the difficulties of the additive  and multiplicative models to model high and low SNR regions, respectively. The interpolated model is able to adapt to both regions, though outperformed by the hierarchical model. The performance gap between the hierarchical model and the other models increases with the multiplicative noise level. This shows that the increase in the multiplicative noise level has a significant impact on the quality of all three approximate methods in the intermediate regime. In this regime, the true likelihood is the farthest from either a Gaussian or a lognormal distribution. 

\begin{table}
    \centering
    \caption{Pairwise comparison of MCMC models based on Exact Expected Log Predictive Density (ELPD) for different multiplicative noise levels ($\sigma_m$). A positive mean $\Delta\textrm{ELPD}\nl^{(i,j)}$ indicates that Model $i$ has higher predictive accuracy than Model $j$.}
    \label{tab:elpd_pairwise_noise}
    \begin{tabular}{@{}ll ccc @{}}
        \toprule
        & & \multicolumn{3}{c}{\textbf{Mean $\Delta\textrm{ELPD}\nl^{(i,j)}$}} \\
        \cmidrule(lr){3-5}
        & & \multicolumn{3}{c}{\textbf{Noise Level} ($\sigma_m$)} \\
        \cmidrule(lr){3-5}
        \textbf{Model $i$} & \textbf{Model $j$} & $\log(1.1)$ & $\log(1.5)$ & $\log(2.0)$ \\
        \midrule
        Hierarchical   & Interpolate    & $0.003$  & $0.059$  & $0.182$ \\
        Hierarchical   & Multiplicative & $0.043$  & $0.099$  & $0.222$ \\
        Hierarchical   & Additive       & $0.004$  & $0.079$  & $0.225$ \\
        Interpolate    & Multiplicative & $0.040$  & $0.040$  & $0.040$ \\
        Interpolate    & Additive       & $0.001$  & $0.019$  & $0.043$ \\
        Multiplicative & Additive       & $-0.039$ & $-0.021$ & $0.003$ \\
        \bottomrule
    \end{tabular}
\end{table}

Finally, the median of the computing time per iteration over all seeds and noise levels is 8.26s for the hierarchical model, 5.34s for the interpolated model, 4.47s for the additive model and 4.63s for the multiplicative model. The hierarchical model maintains a limited computational cost, despite increasing the number of parameters per pixel from $D$ for approximate models to $D + L$. This represents 3.5 times more parameters to be inferred in this realistic astronomical example. 

These results and those presented in the supplementary material show the benefits of the hierarchical likelihood model over approximations. The proposed model is able to accurately model all regions of the parameter maps over a wide range of noise levels. In addition, the hierarchical approach alleviates the need to calibrate additional hyperparameters in an approximate likelihood function, which can be costly and tedious in practice. This makes the proposed approach robust and versatile, while providing good performance over all the metrics considered.

\section{Conclusion}
\label{sec:conclusion}

This work addresses inverse problems subject to additive and multiplicative sources of noise as well as censoring and a highly non-linear forward model all at once. Such problems induce a potentially multimodal posterior distribution. 
To this end, the proposed hierarchical Bayesian model avoids approximations of the likelihood that could be either oversimplistic or cumbersome to fine-tune. 
This is a major advantage in real world applications where no ground truth is available. 
It is also very general, as it can account for other sources of noise by introducing new levels in the hierarchy of the model.
An efficient MCMC algorithm permits to form reliable estimates with quantified uncertainties.
More specifically, an efficient mechanism enables to sample the auxiliary variable introduced by the hierarchical approach, leading to a limited computational overhead compared to the state-of-the-art.

Performances have been evaluated on an inverse problem in astronomy combining all the challenges mentioned above.
The proposed sampler provides similar performance to our revised version of~\cite{Palud_Thouvenin_Chainais_Bron_LePetit_2023} (see supplementary material, Section 2) in terms of reconstruction quality, but with significantly less hyperparameters to fine-tune. Another significant improvement is observed in terms of the ELPD metric: it shows the better predictive performance of the hierarchical model. It also outlines the better general relevance and versatility of the proposed hierarchical model over a wide range of noise and censoring levels. The resulting quantification of uncertainty is therefore more reliable. More importantly, this paper provides a comprehensive model study that will help practitioners to make an informed decision about the most suitable likelihood model with respect to specific application contexts. A cherrypicked simple approximate approach might be sufficient for some noise levels when point-wise estimates are the only concern, but the proposed hierarchical model is more relevant when the robustness of predictive performance and guarantees on uncertainty quantification are crucial.

A natural perspective is the application of the proposed hierarchical model to real astronomical data~\cite{Palud_Bron_Chainais_Petit_Thouvenin_Santa-Maria_Goicoechea_Languignon_Gerin_Pety_et_al_2025}. To scale to the higher-dimensional problems often encountered in such applications, future works can address the design of a more efficient proposal in the I-MTM transition kernel. Indeed, maintaining multiple candidates can be computationally very intensive in higher dimensions. This bottleneck could benefit from a distributed implementation, for which the hierarchical model and the associated MCMC sampler are naturally well suited.

\section*{Acknowledgements}
The authors warmly thank Dr. Pierre Palud and members of the \href{https://www.iram.fr/~pety/ORION-B/}{ORION-B consortium} for interesting discussions at an early stage of this project.

% \printbibliography

% Column equalization (from http://texdoc.net/texmf-dist/doc/latex/IEEEtran/IEEEtran_HOWTO.pdf)
% \IEEEtriggeratref{23} % equalize columns starting from entry 12 in the biblio
% \IEEEtriggercmd{\enlargethispage{-5.35in}} % avoid newpage default after the above command
\bibliographystyle{IEEEtran}
\bibliography{biblio}

\appendices

\section{\texorpdfstring{Fitted Gamma distributions for $q_u$}{Fitted Gamma distributions for proposal on u\_nl}}
\label{sec:fitting_procedure_proposal_u}
At each iteration $t$ of the sampler, the parameters of the Gamma proposal distribution $q_u$ used for non-censored observations are adjusted to fit the mode and the curvature of the conditional distribution~\eqref{eq:cond_u}.
For readability, the iteration index $t$ is omitted in the following.
The mode $u\nl^{\star}$ of \eqref{eq:cond_u} is found by minimizing the non-convex function
\begin{equation}
  \textsf{f}(u\nl) = - \log \pi(u\nl \mid y\nl, \tilde{\btheta}_n),
  \label{eq:appendix_objective}
\end{equation} 
using a projected Newton-Raphson optimization scheme \cite{Nocedal_Wright_2006}.
However, a random initialization can fall in very low probability regions of the conditional distribution (\ref{eq:cond_u}) due to its high dynamic range. 
This can prevent the Newton-Raphson algorithm from converging towards the mode $u\nl^{\star}$.

The initialization $u\nl^{(0)}$ should therefore be carefully chosen to ensure convergence. The mode $u\nl^{\star}$ is further known to lie in the interval between $m_m=f_{\ell}(\tilde{\btheta}_n)\exp(-\frac{3\sigma_m^2}{2})$ and $m_a=y\nl$, the modes of $\pi(u\nl \mid \tilde{\btheta}_n)$ and $\pi(y\nl \mid u\nl)$, respectively.
The initialization $u\nl^{(0)}$ is determined by first evaluating \eqref{eq:appendix_objective} on a coarse 1D grid $\boldsymbol{g}\in(\mathbb{R}_+^*)^P$ of $P$ points evenly spaced in log scale in the interval $[\min(m_m, m_a), \max(m_m, m_a)]$. We assume that $u\nl^{\star}$ lies between the highest-density grid point and its adjacent neighbor with the larger density. These two consecutive points might provide a wide interval around the true mode because of the coarse resolution of the grid. The initialization point is then defined using a weighted mean of this selected pair of points, indexed by $i'$ and $i'+1$, and based on the inverse of the first-order derivative of $\textsf{f}$
\begin{equation}
  u\nl^{(0)} = \frac{g_{i'} \,s(g_{i'}) \: + \: g_{i'+1} \,s(g_{i'+1})}{s(g_{i'}) + s(g_{i'+1})},
  \label{eq:init_u}
\end{equation}
where $s(g) = \lvert \frac{\partial}{\partial g} \textsf{f}(g) \rvert^{-1}$ is the weight function. This choice is motivated by the fact that smaller gradients are associated with points closer to the mode.

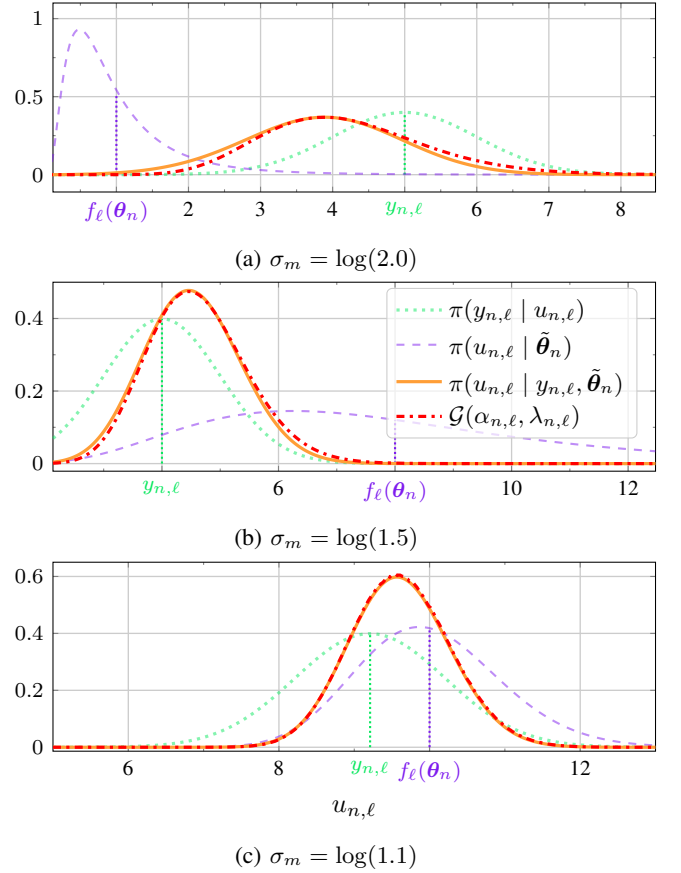
\begin{figure}
% (a) Low Regime
\subfloat[$\sigma_m = \log(2.0)$\label{sfig:low_regime}]{
  
    \begin{tikzpicture}
    \begin{axis}[
        my common style,
        domain=0.01:8.48,
        xmin=0.12, xmax=8.48, ymin=-0.111, ymax=1.1,
        xtick={0,1,2,3,4,5,6,7,8,9},
        xticklabels={0,\empty,2,3,4,\empty,6,7,8,9},
        extra x ticks={1.0, 5.0},
        extra x tick labels={\textcolor{mypurple}{$f_\ell(\btheta_n)$}, \textcolor{mycyan}{$y\nl$}},
        ytick={-0.5,0.0,0.5,1.0,1.5,2.0,2.5}
    ]
    
    \addplot[mycyan,very thick,dotted, opacity=0.5]
        {1/(sqrt(2*pi*1.0^2)) * exp(-0.5*((x-5)/1.0)^2)};
    \draw[mycyan, thick, densely dotted] (axis cs:5,0) -- (axis cs:5,0.399);

    \addplot[mypurple,thick,dashed, opacity=0.5]
        {1/(x*0.6931471805599453*sqrt(2*pi)) * exp(- (ln(x) - (-0.2402265069591007))^2 / (2*0.6931471805599453^2))};
    \draw[mypurple, thick, densely dotted] (axis cs:1.0,0) -- (axis cs:1.0,0.52);

    \addplot[myorange,very thick]
        {((1/(sqrt(2*pi*1.0^2)) * exp(-0.5*((x-5)/1.0)^2)) *
        (1/(x*0.6931471805599453*sqrt(2*pi)) * exp(- (ln(x) - (-0.2402265069591007))^2 / (2*0.6931471805599453^2)))) / 0.006062775375000879};

    \pgfmathsetmacro{\klow}{13.865456279401082}
    \pgfmathsetmacro{\thetalow}{0.3021382834984641}
    \pgfmathsetmacro{\loggammaapproxlow}{0.5*ln(2*pi) + (\klow-0.5)*ln(\klow) - \klow}
    \addplot[red,very thick,dashdotted]
        {exp((\klow-1)*ln(x) - x/\thetalow - \klow*ln(\thetalow) - \loggammaapproxlow)};
        
    \end{axis}
    \end{tikzpicture}
}

\vspace{0.1cm} 

% \hspace{-2em}
% (b) Medium Regime

\subfloat[$\sigma_m = \log(1.5)$\label{sfig:medium_regime}]{
    \begin{tikzpicture}
    \begin{axis}[
        my common style,
        domain=2.13:12.47,
        xmin=2.13, xmax=12.47, ymin=-0.02096, ymax=0.5,
        xtick={2,4,6,8,10,12,14},
        xticklabels={2,\empty,6,\empty,10,12,14},
        extra x ticks={4.0, 8.0},
        extra x tick labels={\textcolor{mycyan}{$y\nl$}, \textcolor{mypurple}{$f_\ell(\btheta_n)$}},
        ytick={0, 0.2, 0.4}
    ]
    
    \addplot[mycyan,very thick,dotted,opacity=0.5] 
        {1/(sqrt(2*pi*1.0^2)) * exp(-0.5*((x-4)/1.0)^2)};
    \draw[mycyan, thick, densely dotted] (axis cs:4,0) -- (axis cs:4,0.399);
    \addlegendentry{$\pi(y\nl\mid u\nl)$}

    \addplot[mypurple,thick,dashed,opacity=0.5] 
        {1/(x*0.4054651081081644*sqrt(2*pi)) * exp(- (ln(x) - 1.997240564733253)^2 / (2*0.4054651081081644^2))};
    \draw[mypurple, thick, densely dotted] (axis cs:8.0,0) -- (axis cs:8.0,0.120);
    \addlegendentry{$\pi(u\nl\mid \tilde{\btheta}_n)$}

    \addplot[myorange,very thick] 
        {(1/(sqrt(2*pi*1.0^2)) * exp(-0.5*((x-4)/1.0)^2)) * (1/(x*0.4054651081081644*sqrt(2*pi)) * exp(- (ln(x) - 1.997240564733253)^2 / (2*0.4054651081081644^2))) / 0.07705217893887424};
    \addlegendentry{$\pi(u\nl \mid y\nl, \tilde{\btheta}_n)$}

    \pgfmathsetmacro{\kmed}{29.031249019577608}
    \pgfmathsetmacro{\thetamed}{0.15912205258676862}
    \pgfmathsetmacro{\loggammaapproxmed}{0.5*ln(2*pi) + (\kmed-0.5)*ln(\kmed) - \kmed}
    \addplot[red,very thick,dashdotted] 
        {exp((\kmed-1)*ln(x) - x/\thetamed - \kmed*ln(\thetamed) - \loggammaapproxmed)};
    \addlegendentry{$\mathcal{G}(\alpha\nl, \lambda\nl)$}

    \end{axis}
    \end{tikzpicture}
}

\vspace{0.1cm} 
% (c) High Regime
\subfloat[$\sigma_m = \log(1.1)$\label{sfig:high_regime}]{
    \begin{tikzpicture}
    \begin{axis}[
        my common style,
        domain=5:13,
        xmin=5, xmax=13, ymin=-0.01385, ymax=0.65,
        xtick={6,8,10,12},
        xticklabels={6,8,\empty,12},
        extra x ticks={9.2104, 10.0},
        extra x tick labels={\textcolor{mycyan}{$y\nl$}, \textcolor{mypurple}{$f_\ell(\btheta_n)$}},
        ytick={-0.1,0,0.2,0.4,0.6,0.8,1.0},
        xlabel={$u\nl$}
    ]
    
    \addplot[mycyan,very thick,dotted,opacity=.5] 
        {1/(sqrt(2*pi*1.0^2)) * exp(-0.5*((x-9.210408641317397)/1.0)^2)};
    \draw[mycyan, thick, densely dotted] (axis cs:9.2104,0) -- (axis cs:9.2104,0.399);

    \addplot[mypurple,thick,dashed,opacity=.5] 
        {1/(x*0.09531017980432495*sqrt(2*pi)) * exp(- (ln(x) - 2.2980430778068794)^2 / (2*0.09531017980432495^2))};
    \draw[mypurple, thick, densely dotted] (axis cs:10.0,0) -- (axis cs:10.0,0.418);

    \addplot[myorange,very thick] 
        {(1/(sqrt(2*pi*1.0^2)) * exp(-0.5*((x-9.210408641317397)/1.0)^2)) * (1/(x*0.09531017980432495*sqrt(2*pi)) * exp(- (ln(x) - 2.2980430778068794)^2 / (2*0.09531017980432495^2))) / 0.2512574916545743};

    \pgfmathsetmacro{\khigh}{205.96992412520135}
    \pgfmathsetmacro{\thetahigh}{0.046666346478558746}
    \pgfmathsetmacro{\loggammaapproxhigh}{0.5*ln(2*pi) + (\khigh-0.5)*ln(\khigh) - \khigh}
    \addplot[red,very thick,dashdotted] 
        {exp((\khigh-1)*ln(x) - x/\thetahigh - \khigh*ln(\thetahigh) - \loggammaapproxhigh)};

    \end{axis}
    \end{tikzpicture}
}

\caption{Examples of fitted Gamma proposal distributions (red dotted line) for different cases.}
\label{fig:fitting-gamma-examples}
\end{figure}

A projected Newton-Raphson optimization algorithm is then used to find an estimate of $u\nl^\star$ within $M$ iterations. 
Matching the estimated mode $u\nl^{(M)}$ and the second-order derivative of $\mathsf{f}$ with those of the Gamma distribution at the point $u\nl^{(M)}$ leads to closed-form expressions for the shape parameter $\alpha\nl$ and the rate parameter $\lambda\nl$ of the Gamma proposal distribution. 
This complete procedure is summarized in \Cref{alg:fitting-gamma}.
\begin{algorithm}
\caption{Fitting procedure of the Gamma proposal distribution for uncensored $u\nl$.}
\label{alg:fitting-gamma}xr
\begin{algorithmic}[1]
\State \textbf{Input:} $f_\ell(\tilde{\btheta}_n)$, $y\nl$, $\boldsymbol{\sigma}_m$, $\boldsymbol{\sigma}_a$, $M$, $P$
\State $m_m = f_\ell(\tilde{\btheta}_n)\exp(-\frac{3\sigma_{m}^2}{2})$, $m_a = y\nl$ 
\State $m_l = \min(m_a, m_m)$, $m_u = \max(m_a, m_m)$
\State $\triangleright$ Logarithmically spaced grid of $P$ points in $[m_l, m_u]$
\State $\boldsymbol{g} = (g_i)_{i=1}^P$ with $g_i = m_l \left( \frac{m_u}{m_l} \right)^{\frac{i-1}{P-1}}$ 
\State $i' =  \underset{i\in \llbracket 1, P-1 \rrbracket}{\arg\,\min} \, \big(\textsf{f}(g_i) + \textsf{f}(g_{i+1}) \big)$
\State Set $u\nl^{(0)}$ using \eqref{eq:init_u}
\State $\triangleright$ Projected Newton-Raphson
\For{$i = 1$ to $M$}
  \State $u\nl^{(i)} = u\nl^{(i-1)} - \frac{\partial}{\partial u\nl} \textsf{f}(u\nl^{(i-1)}) / \lvert \frac{\partial^2}{\partial {u\nl}^2} \textsf{f}(u\nl^{(i-1)}) \rvert$
  \State $u\nl^{(i)} = \min\bigl(m_u, \,\max(m_l, \, u\nl^{(i)})\bigr)$
\EndFor
\State $\alpha\nl = 1 + {u\nl^{(M)}}^2 \lvert \frac{\partial^2}{\partial {u\nl}^2} \textsf{f}(u\nl^{(M)}) \rvert$
\State $\lambda\nl = \frac{\alpha\nl -1}{u\nl^{(M)}}$
\State \textbf{Output:} Gamma proposal $\mathcal{G}(\alpha\nl, \lambda\nl)$ for $u\nl$
\end{algorithmic}
\end{algorithm}
Values of $M=5$ and $P=10$ yielded satisfactory results for all the experiments described in \Cref{sec:experiments}.
\Cref{fig:fitting-gamma-examples} shows examples of fitted Gamma proposal distributions for different values of $f_\ell(\tilde{\btheta}_n)$ and $y\nl$. 

\end{document}